\documentclass[twocolumn,
aps,
prb,
amsmath,
amssymb,
floatfix,
]{revtex4-1}

\pdfoutput=1

\usepackage{xcolor,mathrsfs,dsfont}
\definecolor{darkblue}{RGB}{0,0,150}
\definecolor{nightblue}{RGB}{0,0,100}
\definecolor{evergreen}{RGB}{0,120,0}

\usepackage{graphicx,mathtools,bm}
\usepackage[
colorlinks,
citecolor=darkblue,
linkcolor=darkblue,
urlcolor=nightblue]{hyperref}

\pdfoutput=1

\usepackage[english]{babel}
\usepackage[babel,kerning=true,spacing=true]{microtype}
\usepackage{tabulary}

\usepackage{appendix,stmaryrd}

\newcommand{\dd}{\mathrm{d}}
\renewcommand{\Im}{\ensuremath{\mathrm{Im}\,}}
\renewcommand{\Re}{\ensuremath{\mathrm{Re}\,}}

\newcommand{\beq}{\begin{equation}}
\newcommand{\eeq}{\end{equation}}

\bibpunct{[}{]}{,}{n}{}{}
\makeatletter
\def\NAT@def@citea{\def\@citea{\NAT@separator}}
\makeatother

\def\dd{\;\!\mathrm{d}} % differential for integration 

\begin{document}

\title{
Consequences of Time-reversal-symmetry Breaking in the Light-Matter Interaction: Berry Curvature, Quantum Metric and Diabatic Motion
}
\author{Tobias Holder}
\email{tobias.holder@weizmann.ac.il}
\author{Daniel Kaplan}
\author{Binghai Yan}
\email{binghai.yan@weizmann.ac.il}
\affiliation{Department of Condensed Matter Physics,
Weizmann Institute of Science,
Rehovot 7610001, Israel}
\date{\today}

\begin{abstract}
Nonlinear optical response is well studied in the context of semiconductors and has gained a renaissance in studies of topological materials in the recent decade. So far it mainly deals with non-magnetic materials and it is believed to root in the Berry curvature of the material band structure. In this work, we revisit the general formalism for the second-order optical response and focus on the consequences of the time-reversal-symmetry ($\mathcal{T}$) breaking, by a diagrammatic approach. We have identified three physical mechanisms to generate a dc photocurrent, i.e. the Berry curvature, a term closely related to the quantum metric, and the diabatic motion. All three effects can be understood intuitively from the anomalous acceleration. The first two terms are respectively the antisymmetric and symmetric parts of the quantum geometric tensor. The last term is due to the dynamical antilocalization that appears from the phase accumulation between time-reversed fermion loops. Additionally, we derive the semiclassical conductivity that includes both intra- and interband effects. 
We find that $\mathcal{T}$-breaking can lead to a greatly enhanced non-linear anomalous Hall effect that is beyond the contribution by the Berry curvature dipole.
\end{abstract}

\maketitle

%(21,23,38,39)

\section{Introduction}

The Bulk photovoltaic effect (BPVE)~\cite{Belinicher1980,vonBaltz1981,Boyd2003} refers to the generation of a dc current from a uniform material by irridation with strong light. Since the early 1980s, its understanding has been gradually established as the second-order optical response~\cite{Kraut1979,vonBaltz1981,Sipe2000,Young2012,Ventura2017,Parker2019} that is closely related to the Berry phase~\cite{Xiao2010}, an intrinsic quantity of the material band structure. Recently it gained renewed interest in topological Weyl semimetals (WSMs)~\cite{Yan2017,Armitage2017}, where the Berry phase is believed to generate the giant photocurrent and also the second harmonic response~\cite{Wu2017,Ma2017,osterhoudt2019colossal,Hosur2013,Morimoto2016,Taguchi2016,Chan2017,deJuan2017,Zhang2018}. 
Thus far, research on BPVE focuses on nonmagnetic materials. However, recent theoretical~\cite{Zhang2019} and experimental~\cite{Ganichev2002,Ganichev2006,Sun2019} works on magnetic systems reveal a distinct photocurrent and second harmonic generation which cannot be merely derived from the Berry phase. Therefore, we are motivated to re-examine the second-order response theory and investigate the effects of time-reversal-symmetry ($\mathcal{T}$) breaking.

\subsection{Overview}
The non-linear current $\bm{j}$ created at the second order in the incident electric field $\bm{E}$ of light is defined as 
\begin{align}
    j^c(\omega;\omega_1,\omega_2)&=
    \sum_{ab}\sigma^{ab;c}(\omega;\omega_1,\omega_2)E^a(\omega_1)E^b(\omega_2)
\end{align}
A focal point will be in the following the dc-current $\bm{j}(0;\omega,-\omega)$ which is created in response to irradiation of light with finite frequency $\omega$.
In a clean, gapped system with $\mathcal{T}$, the BPVE is usually phrased in the form of two phenomena, respectively termed  shift current and injection current. Both effects require inversion symmetry breaking, otherwise the momentum-space integral of the response function vanishes. The shift current denotes the current that is proportional to the positional shift (shift vector expressed in the form of the Berry connection) of the electron charge in an interband process. The shift current has previously been employed to successfully describe the BPVE in many compounds
~\cite{vonBaltz1981,Young2012,Young2012a,Morimoto2016a,Cook2017,Tan2016}.
Injection current is a stronger response, which is described by a constant current creation with rate $\dd \bm{j}/\dd t$~\cite{Sipe2000,Olbrich2009,Yuan2014,McIver2012,Okada2016,Koenig2017,Golub2018,deJuan2019}. In systems with finite relaxation rates, dissipative processes will limit the current injection, thus rendering the nonlinear conductivity finite, with a magnitude proportional to the lifetime of the quasiparticles. In materials with $\mathcal{T}$, the injection current will occur only in response to circular polarized light and a linear polarization can only lead to a shift current. This distinction is lost in compounds which break $\mathcal{T}$, as pointed out by Ref.~\onlinecite{Zhang2019} very recently, where both shift and injection current might arise for either linear or circular polarized light. 
While the non-linear currents created by the incident light usually lead to Joule heating in the sample, it is possible in experiment to subtract this effect. In accordance with previous literature, we therefore present results only for the electrical conductivities, and not for thermoelectric response.

Compared to the shift current generation, the intuitive understanding of current injection is less clear. One of the initial objectives of this work was to understand its physical origin. An important step was the realization that a single cone in a two-band Weyl model can produce injection current at a quantized rate, which is proportional to the topological charge of a Weyl cone~\cite{deJuan2017}. 
However, it seems unlikely for this identification of the injection current with topological charge to translate to the generic case. Also, an analogous procedure to express the shift current with the Berry curvature has not been reported.
Given these findings, it is presently unclear to which extent the presence of Berry monopoles located at the Weyl nodes determine the BPVE in  WSMs. Indeed, topologically trivial semimetals like Graphene have zero Berry curvature and are still known to possess a whole assortment of unique transport properties, among them an abnormally large third order conductivity, merely due to their nodal points in the band structure and the surrounding linear Dirac cone dispersion~\cite{CastroNeto2009,Chen2014,Chen2015}. 
A positive statement about the topological origin of the large shift current observed for example in TaAS~\cite{Wu2017} thus requires an identification of this response with an observable related to the topological nature of the material. 

Recently, an intriguing nonlinear anomalous Hall effect (AHE)~\cite{Deyo2009,Moore2010,Sodemann2015,Low2015,Tsirkin2018} was derived in the semiclassical limit, which refers to intraband transitions in the small frequency case of the BPVE. It is attributed to the existence of the Berry curvature dipole in the semiclassical description of the second-order response~\cite{Sodemann2015,Morimoto2016} and was shortly after discovered in thin films of the WSM WTe$_2$~\cite{Zhang2018a,Zhang2018b,Ma2019,Kang2019}. It is, however, unclear how the mechanisms for shift and injection currents, which are from interband transitions, are related to the semiclassical motion in terms of the Berry curvature dipole.

The present discussion also touches upon a practical question of how to properly model nonlinear response in realistic quantum systems with impurities or at finite temperature. 
For example, there is a longstanding question how the canonical perturbation theory is related to results from the kinetic approach~\cite{Belinicher1980,Sturman1992,Ivchenko1997,Ivchenko2005}. In the latter approach, the current response to linear polarized light features not only a shift current but also a ballistic current\cite{Sturman2019}, which originates from the momentum space asymmetry of the distribution function. In particular, in the derivation of the ballistic current, the transport time enters twice, once in the numerator to account for the build-up of the asymmetry in the distribution function, and once in the denominator as a result of phonon corrections to the electron-photon vertex. As a result, the ballistic current emerges as an order-one effect which appears completely unaccounted for in an adiabatic treatment.

In perturbation theory, a large part of the discussion has been framed in the context of two gauge choices, the length gauge ~\cite{Sipe2000,Young2012}, and the velocity gauge~\cite{Passos2018,Parker2019}. These names are derived from the way the applied electric field is coupled to the Hamiltonian. In the length gauge, electric field enters via a dipole energy which induces a polarization, while in the velocity gauge it is included through the electromagnetic gauge potential via minimal substitution in the momentum. Both approaches were originally conceived as complementary descriptions, but in the clean limit they yield identical results and can be related by a time-dependent gauge transformation~\cite{Ventura2017,Taghizadeh2017}. 
Of course, also for finite quasiparticle lifetimes the calculated optical response is always independent of the chosen gauge. However, depending on the choice of gauging, different choices for the relaxation rates have been proposed in the respective formulation~\cite{Mikhailov2016,Passos2018}. 
The issue is that non-linear optical response fundamentally involves interband processes which are difficult to model semiclassically. As such, there is some ambiguity with regards to the lifetimes which enter the expressions. 
While we will not attempt to solve this issue comprehensively, we offer a way to discuss both the adiabatic and the diabatic motion which arises under optical driving in connection with the so-called quantum geometric tensor~\cite{Kolodrubetz2017}. Since this object has a clear physical interpretation, it is possible to retrace which finite relaxation rates enter the expressions.
Here, we focus on the intrinsic mechanism. We leave the discussion of extrinsic mechanisms for a future work. For time-reversal invariant systems, this has been worked out recently~\cite{Isobe2018,Du2019,Koenig2019}.
We emphasize that scattering induced processes can be comparable in size to the response due to band structure effects~\cite{Deyo2009,Sturman2019}.

In the following, we proceed to systematically explore the second-order optical response where not only inversion symmetry is absent but also the $\mathcal{T}$ is broken. In particular, we derive formulas for the semiclassical response, shift current and injection current in terms of observable, physically distinct processes, some of which lead to previously unknown types of photocurrents as long as $\mathcal{T}$ is absent.
Our starting point is the treatment of the BPVE in the diagrammatic approach~\cite{Parker2019} which can also account for finite lifetimes and offers a more immediate interpretation of the different contributions to second order response. 
Still, these expressions remain opaque to a semiclassical interpretation of the physical processes as described by the perturbation theory.
Therefore, we propose to understand the nonlinear response by rewriting the second-order conductivity in terms of semiclassical accelerations, separately considering instantaneous and sequential multi-photon processes, as dictated by the diagrammatics. 

\subsection{Short Summary of Results}
As the main results, we demonstrate that the second-order optical response can be understood in terms of three distinct physical processes, each of them a part of the matrix elements of the anomalous quasiparticle acceleration in a Bloch band. 
While it is well known that the velocity matrix elements in band eigenbasis are not simply derivatives of the eigenenergies but contain an anomalous velocity originating from the nonzero Berry curvature\cite{Niu1996}, the anomalous acceleration has not yet been discussed at this same level of detail~\cite{Sundaram1999,Morimoto2016a,Matsyshyn2019,Gao2019}. 
To be precise, consider a dispersion $\hbar\varepsilon_m$ with band index $m$ and eigenstate vectors $|m\rangle$, its momentum derivative $v^a_{mn}$ for the spatial direction $a$ and the Berry connection $r^a_{mn}=\langle m|\partial_{k_a}|n\rangle$.
The semiclassical center of mass motion of a wavepacket with position $\bm{r}_{CM}$ and momentum $\bm{k}$ moving in band $m$ subject to an electric field $\bm{E}$ has the equation of motion
\begin{align}
    \dot r^a_{CM}&=
    \sum_m\partial_{k_a} \varepsilon_m
    -\tfrac{e}{\hbar}E^b\Omega^{ba}_{mm},
    \label{eq:anomvelocity}
\end{align}
where the linear-response form of the anomalous velocity $\tfrac{e}{\hbar}E^b\Omega^{ba}_{mm}$ is due to the Berry curvature $\Omega^{ba}_{mm}=\partial^br^a_{mm}-\partial^ar^b_{mm}$ which is usually written in the form of a vector $\Omega^c=\epsilon^{abc}\Omega^{ab}_{mm}$.
As a reminder, this result is most easily derived in the adiabatic approach by employing the identification $\partial_t=(-e/\hbar) E^a\partial_{k_a}$, which follows from the semiclassical Lagrangian in minimal coupling~\cite{Xiao2010}. This means that the acceleration can be found by considering the expectation value $\ddot r^a_{CM}=-E^b\partial_{k_b}\dot r^a_{CM}$. However, this derivation of the acceleration is problematic if  bilinear couplings to the electric field are to be retained, as this requires the equations for the semiclassical motion to be correct up to second order.
Indeed, \citeauthor{Gao2014}~\cite{Gao2014} reported some time ago that to second order in applied fields, there appears not only a polarization dependent renormalization of the dispersion $\tilde\varepsilon(\bm{E})$, but also a field dependent positional shift $\bm{a'}(\bm{E})$ of the semiclassical wave packet, both of which enter the equation of motion for a wavepacket,
\begin{align}
    \bm{\dot r}_{CM}&=\nabla\tilde\varepsilon-\bm{\dot k}\times (\bm{\Omega}+\nabla\times\bm{a'}).
\end{align}
However, the discussion at the time was mostly concerned with the AHE and did not resolve these modifications with respect to second order optical response. 
We remedy this situation, but for the discussion at hand it turns out to be better to directly consider the matrix elements of the second momentum derivative $w^{ab}_{mn}=\langle m |\partial_{k_a}\partial_{k_b} H_0(\bm{k})|n\rangle$ of the Hamiltonian $H_0(\bm{k})$, 
\begin{align}
    w^{ab}_{mn}
    &=
    i\sum_l(v^a_{ml}r^b_{ln}-r^b_{ml}v^a_{ln})
    \notag\\&\quad
    +i (v^a_{mm}-v^a_{nn})r^b_{mn}
    \notag\\&\quad
    +i(\varepsilon_{m}-\varepsilon_{n}) R^{ab}_{mn}
    \notag\\&\quad
    +\delta_{mn}\partial_{k_a}\partial_{k_b}\varepsilon_m,
    \label{eq:anomacc}
\end{align}
where the symbol $R^{ab}_{mn}=(\langle i \partial_{k_a}\partial_{k_b} m| n\rangle+\langle m| i \partial_{k_a}\partial_{k_b} n\rangle)/2$ will be explained momentarily.
In Eq~\eqref{eq:anomacc}, the last term is the normal derivative (i.e. a Drude term), while the other three give rise to the anomalous acceleration.
These features of the anomalous motion appear in the second order optical response due to the close connection between the electric field and the derivative in time. However, a straightforward translation between anomalous motion and optical response is complicated by the fact that the perturbative response formulas intrinsically contain interband processes and thus off-diagonal matrix elements in the band basis. For this reason it went so far unnoticed that second-order response actually contains three types of anomalous contributions, unlike the linear response, where only the Berry curvature makes an appearance. The phenomenology is summarized in Tab.~\ref{tab:I}.

(i) The first term in Eq~\eqref{eq:anomacc} is the well-studied adiabatic acceleration of the electron wavefunction due to the dipole interaction with the electric field. In the second-order response, this can always be formulated as a properly symmetrized triple product of velocity matrix elements, and it corresponds to the Berry curvature dipole term in the simplified semiclassical treatment. Importantly, this adiabatic acceleration appears in the shift current.

(ii) The second term is a diabatic effect, which can also be written as a velocity difference between occupied and unoccupied bands. This response type can be identified as a mismatch between the derivative $\partial_{k_j} v^i$ of the quasiparticle velocity $v^i$ and the derivative $\varepsilon\partial_{k_j} r^i$ of the Berry connection $r^i$. Speaking in terms of Feynman diagrams, the phases accumulated on momentum reversed (i.e. time reversed) paths do not cancel with each other. This phenomenon is responsible for current injection, which can therefore be viewed as the result of \emph{dynamical}, i.e. finite-frequency antilocalization between time-reversed loops with three legs (see Fig.~\ref{fig:diags}).
We also show that remnants of the diabatic motion appear in the shift current as a projective correction to the Berry dipole. This correction is reminiscent of the ballistic current as found in the kinetic approach.

(iii) The third term resembles what is known as the bare quantum metric. Somewhat surprisingly, it is impossible to write higher order conductivities only as a function of velocity matrix elements and the Berry curvature. 
This is manifest in the third type of anomalous acceleration, which is constituted by the symmetric part of the second derivative of the wave function,
\begin{align}
    R^{ab}_{mn}=\tfrac{1}{2}(\langle i \partial_{k_a}\partial_{k_b} m| n\rangle+\langle m| i \partial_{k_a}\partial_{k_b} n\rangle), \end{align}
which is closely related but not identical to the gauge-invariant quantum metric $g^{ab}$~\cite{Anandan1990,Ma2010,Cheng2010,Kolodrubetz2017,Ozawa2018,Zhang2019a}. 
In essence, second order response contains information about both the symmetric  and antisymmetric part of the quantum geometric tensor (QGT),
\begin{align}
    Q^{ab}_{mn}&=\langle \partial^a m|1-\mathcal{P}_{GS}|\partial^b n\rangle\notag\\
&=g^{ab}_{mn}-\tfrac{i}{2}\Omega^{ab}_{mn},
\end{align}
where $\mathcal{P}_{GS}$ is the projector onto the ground state manifold.
In comparison, the linear response coefficients receive a contribution from the Berry curvature $\Omega^{ab}_{mm}$ only. 
We emphasize that both the Berry curvature and the bare quantum metric enter the second-order conductivity not directly, but in a form resembling a momentum derivative of these quantities. This is also the reason why it is necessary to consider the off-diagonal matrix elements of the acceleration, and why it is helpful to intermittently consider gauge-covariant and not exclusively gauge-invariant quantities. 
This nonwithstanding, our final expressions for the different contributions to the non-linear conductivity are gauge-invariant.

\begin{table}
    \begin{tabular}{p{2.4cm} p{1.5cm} p{1.4cm} p{1.5cm} p{1.cm}}
    \hline
     & Berry curvature& Diabatic & Quantum Metric & Drude \\
    \hline
    Shift current    & \checkmark & & \checkmark &\\
    Injection current & & \checkmark & &\\
    Leading-SC     & (\checkmark) & (\checkmark) &  & (\checkmark) \\
    Next-leading-SC    & \checkmark & (\checkmark) & &\\
    \hline
    \end{tabular}
    \caption{Overview of different contributions to second-order response in terms of acceleration processes. The four sources of nonlinear current are the Berry-type geometric contribution, the diabatic motion, the effect of the quantum metric and the Drude-type term (density of states). Checkmarks in brackets denote the new terms which only appear when the material lacks $\mathcal{T}$. The leading order term (leading-SC) and next-leading order term (next-leading-SC) represent the semiclassical (SC) expansion at finite frequency.}
    \label{tab:I}
\end{table}

(iv) Finally, we derive the semiclassical conductivity in the absence of $\mathcal{T}$ by employing a canonical expansion of the conductivity in powers of the applied frequency ($\omega$). The leading order term ($\omega^{-2}$) appears only when breaking $\mathcal{T}$ and contains three of the above four contributions, as summarized in Table~\ref{tab:I}. We find that nonlinear response contains terms of geometric origin exclusively in the next-leading order in frequency ($\omega^{-1}$). As $\mathcal{T}$ is broken, this next-leading order term is not given only by the Berry curvature dipole, but contains an additional diabatic term due to the intrinsically dissipative nature of optical transitions. 

In addition, we report a robust way to obtain explicit sum rules for a general tight-binding Hamiltonian required to translate between velocity gauge and length gauge, even if the system is imperfect and quasiparticle lifetimes are finite. 
Our derivation shows that the sum rules should be viewed as an expansion of higher order vertices in terms of velocities and band energies. This is in agreement with the general formulation of sum rules in clean systems, where they appear in terms of vanishing nested covariant derivatives of an observable, as mandated by gauge invariance~\cite{Ventura2017}. However, the latter formulation is entirely formal, leaving the resolution of the derivatives of matrix elements untouched. It is precisely this point where additional lifetime dependencies appear, thus making the sum rules one possible source of disagreement between the length  and the velocity gauge formalism. In particular, the often quoted resolution of the non-Abelian Berry connection in terms of band structure parameters as $r^a_{mn}=v^a_{mn}/i\epsilon_{mn}$ does not hold if the quasiparticle lifetime is finite.
In the past, another point of confusion was the use of commutator identities which are only valid for an infinite number of bands~\cite{Aversa1995,Ventura2017}. For example, the relation
\begin{align}
    [r^a, v^b]=\frac{i\hbar}{m}\delta^{ab}
\end{align}
generally does not hold for a tight-binding Hamiltonian. However, this problem is unrelated to gauging, it originates from the reliance on the relation $ p^a= v^a m$, which is only true for an isotropic dispersion. The diagrammatic formalism does not necessitate assumptions of this kind, therefore the sum rules presented here do not suffer from these limitations.
We find that the sum rules are only effective at reducing the zero frequency singularity of the second-order conductivity if the material is time-reversal symmetric. 

The remaining sections are organized as follows.
In Sec. II, we introduce the general formalism for nonlinear optical response for a clean system, relying mostly on the formulation in the velocity gauge. The introduction of finite lifetimes can be intuitively implemented within the diagrammatic approach. We discuss the phenomenology of this approach and also compare to the equivalent procedure in the length gauge. 
In Sec. III we explain the various processes which contribute to the second order response. Some time is spend on the semiclassical picture associated with current injection, and we derive an intuitive formulation of this phenomenon in terms of a phase difference accumulated on a closed path. We calculate the semiclassical conductivity including all multiband contributions in the absence of $\mathcal{T}$. In Sec. IV, we illustrate our findings by numerical calculations on a concrete example of a WSM and a Dirac semimetal. 
This serves as a model for a system which has zero (Abelian) Berry curvature and yet exhibits a large BPVE. 

\section{Perturbation theory}
We briefly recollect the basic ingredients needed to capture optical response when going beyond linear approximation. In the literature, this was achieved either by calculating the electric susceptibility or more directly by evaluating the current-current correlator to second order in the electric fields. These approaches are also known under the terms length gauge and velocity gauge.

\subsection{Adiabatic formalism}
The current created by the BPVE originates from the intraband motion of Bloch states in response to the irradiation.
The landmark result for this effect was derived in~\cite{Sipe2000} and reads 
\begin{align}
    \sigma^{ab;c}_{dc}(\omega)
    &=\left.\frac{e^3}{\hbar^2 \omega'}\sum_{m,n}\int_{\bm{k}}
    \frac{f_{nm}r^a_{nm}r^b_{mn}\Delta_{mn}^c}{\varepsilon_{mn}-\omega-i0^+}\right|_{\omega'=0}\notag\\
    &-\frac{e^3}{\hbar^2}\sum_{m,n}\int_{\bm{k}}
    \frac{f_{nm}r^a_{mn}r^b_{nm,c}}{\varepsilon_{mn}-\omega-i0^+}
    \notag\\
    &+(a,\omega\leftrightarrow b,-\omega).
    \label{eq:sipegeneral}
\end{align}
Here, $m,n$ are band indices, $\varepsilon_{mn}=\varepsilon_m-\varepsilon_n$ is the difference of the eigenvalues of the unpertubed Hamiltonian $H_0$ in units of frequency, i.e. $\varepsilon_m(\bm{k})=\hbar^{-1}\langle m| H_0(\bm{k}) |m\rangle$; the difference of Fermi factors is $f_{mn}=f_m-f_n$ and finally there is the difference of velocities, $\Delta_{mn}^a=v_{mm}^a-v_{nn}^a$, with $v^a_{mn}=\langle m| \partial_{k_a} H_0(\bm{k}) |n\rangle$.
The non-Abelian Berry connection is $r_{mn}^a=i\langle m|\partial_{k_a} n\rangle$ with the generalized derivative $r^b_{nm,c}=\partial_{k_c} r^b_{nm}-i(r_{nn}^c-r_{mm}^c)r_{nm}^b$, where $r_{nn}^c=i\langle n|\partial_{k_c} n\rangle$ is the Abelian Berry connection. The integration is over the first Brillouin zone, with $\int_{\bm{k}}\equiv \int_{FBZ} \dd^3 k/(2\pi)^3$.
The current created through the bulk photovoltaic effect formally corresponds to an infinite polarizability, i.e. charge imbalances are continuously created. However, the first term in Eq.~\eqref{eq:sipegeneral} for the current is infinite by itself, meaning that the second-order response may even produce an infinite current, a phenomenon termed current injection, i.e. current is continuously created. 
Originally, Eq.~\eqref{eq:sipegeneral} was derived from a  painstaking second order calculation of the adiabatic time evolution of the instantaneous eigenstates, also known as a reduced density matrix formalism.
While this approach is conceptually very clear-cut, it relies on two important factors. On the one hand, the optical processes must be fast enough for lifetime effects to be negligible. This assumption becomes questionable for optical transitions involving resonances with transient states in highly excited levels. In the present context, one would expect the current injection to strongly depend on lifetime effects. Secondly, it is often useful to resolve the derivative $r^b_{nm,c}$ by a sum rule. 
The latter task was recently reexamined in~\cite{Cook2017}, finding for $n\neq m$
\begin{align}
    r^b_{nm,c}&=
    \frac{v^b_{nm}(v^c_{nn}-v^c_{mm})-v^c_{nm}(v^b_{nn}-v^b_{mm})}{-i \varepsilon_{nm}^2}\notag\\
    &+\sum_{p\neq m,n}\left(\frac{v^b_{np}v^c_{pm}}{-i\varepsilon_{nm}\varepsilon_{pm}}
    -\frac{v^c_{np}v^b_{pm}}{-i\varepsilon_{nm}\varepsilon_{np}}\right)+\frac{w_{nm}^{bc}}{i\varepsilon_{nm}}
\label{eq:rmnderiv}
\end{align}
Here, the second derivative $w_{mn}^{ab}=\langle m|\partial_{k_a}\partial_{k_b} H_0(\bm{k})|n\rangle$ makes an appearance. This term is was originally not included because a quadratic Hamiltonian implies $w^{ab}_{nm}=0$ for $n\neq m$. Generically $w$ is only one of several higher order vertices which appear in an order-by-order analysis~\cite{Parker2019}.

Properly including lifetime effects necessitates a closer look at the theoretical underpinnings of the adiabatic formalism. The starting point is a formulation of the problem in the length gauge. Equivalently, it follows from the proposition that the response of the crystal can be approximated by a dipole Hamiltonian~\cite{Sipe2000}
\begin{align}
    H_{dip}(t)&=\sum_{n,m}\int_{\bm{k}}\hbar\bar\varepsilon_{nm}(\bm{k},t)c^\dagger_n(\bm{k})c_m(\bm{k})
\end{align}
where the renormalized dispersion $\bar\varepsilon_{nm}(\bm{k},t)=\varepsilon_n\delta_{nm}-e\hbar^{-1} r_{nm}^a E_a(t)$ contains the Berry connection $r$. While this  construction encapsulates the polarization response in a clean system, it relies on the identification of the adiabatic intraband motion with the motion induced by the electromagnetic field~\cite{Xiao2010}. Namely, the crystal momentum $\bm{q}$ in the isolated system and the gauge invariant momentum $\bm{k}$ are by definition related by $\bm{k}=\bm{q}+e \bm{A}(t)/\hbar$. This implies $\bm{\dot{k}}=-e \bm{E}(t)/\hbar$ under the condition that  $\bm{\dot q}=0$, i.e. if the crystal momentum is conserved. 

\subsection{Diagrammatic formulation}
In the velocity gauge, one instead considers the coupling to the gauge field through minimal substitution of the momentum operator $\bm{k}\to\bm{k}-e\bm{A}/\hbar$ in the unperturbed Hamiltonian $H_0(\bm{k})$.
\begin{figure}
\includegraphics[width=.85\columnwidth]{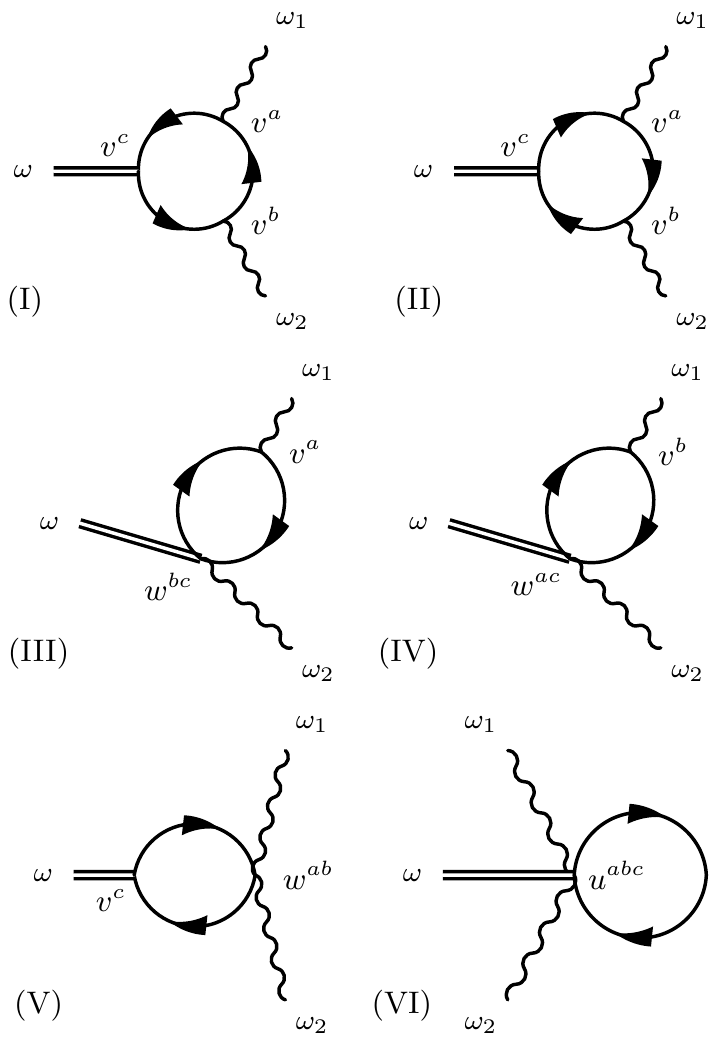}
\caption{Second order contributions to the optical dc-response. Straight lines with arrows denote quasiparticles, a straight double line connects to a current operator and the wavy lines are couplings to the electromagnetic field. Note that diagrams (I) and (II) and (III) and (IV) are variants of each other.}
\label{fig:diags}
\end{figure}
Expanding in powers of $\bm{A}$ yields therefore
\begin{align}
    H_{min}(\bm{k},t)&=H_0(\bm{k})
    +\frac{e}{\hbar}
    \sum_a\frac{\partial H_0(\bm{k})}{\partial k_a} A^a(t)
    \notag\\&\quad
    +\frac{e^2}{2\hbar^2}
    \sum_{a,b}\frac{\partial^2 H_0(\bm{k})}{\partial k_a \partial k_b} A^a(t) A^b(t)
    \notag\\&\quad
    +\frac{e^3}{6\hbar^3}
    \sum_{a,b,c}\frac{\partial^3 H_0(\bm{k})}{\partial k_a \partial k_b \partial k_c} A^a(t) A^b(t) A^c(t)
    \notag\\&\quad
    +\dots
    \label{eq:velocityham}
\end{align}
Here, we wrote everything in terms of operators and not their matrix elements, which would result in a very different expression due to the nature of the (Berry) covariant derivative in k-space, which is defined through the connection
\begin{align}
    \langle m| \partial_{k_a}\mathcal{O}|n \rangle
    &=\frac{\partial \mathcal{O}_{mn}}{\partial_{k_a}}
    -i r^a_{ml}\mathcal{O}_{ln}+i \mathcal{O}_{ml}r^a_{ln}
    \label{eq:connection}
\end{align}
where $\mathcal{O}$ is a Hermitian operator.
This connection plays a major role in the calculation of Eq.~\eqref{eq:velocityham}.
The current operator follows from a variation with respect to the electromagnetic field,
\begin{align}
    J^c&=\frac{\delta H_{min}}{\delta A^c}.
\end{align}
Henceforth we will refer to derivatives in k-space by $\partial^a=\partial/\partial_{k_a}$. The various matrix elements are therefore $v^a_{mn}=\langle m| \partial^a H_0(\bm{k}) |n\rangle$, $w_{mn}^{ab}=\langle m|\partial^a\partial^b H_0(\bm{k})|n\rangle$ and $\langle m|\partial^a\partial^b\partial^c H_0(\bm{k})|n\rangle
=u_{mn}^{abc}$.
We note that using the connection, Eq.~\eqref{eq:connection}, the velocity operator $v^a=\partial^a H_0(\bm{k})$ can be expanded in the eigenbasis to yield the familiar expressions for the anomalous velocity including the Berry curvature term. The same holds for all higher order derivatives. We will defer this step until the discussion of the semiclassical motion.

In total, in second order the conductivity receives contributions from four types of diagrams, two of which have two variants. They are depicted in Fig.~\ref{fig:diags}.
Diagrams (I) and (II) corresponds to the type of response originally considered in the earliest works~\cite{Kraut1979,vonBaltz1981}. Importantly, they only carry velocity matrix elements.
Diagrams (III)-(VI) contain higher derivatives in $\bm{k}$, which in the original formulation of Eq.~\eqref{eq:sipegeneral} appear through $\partial^c r^b_{nm}$, via sum rules.
In the length gauge these higher order derivatives follow from an expansion of the time evolution of the instantaneous eigenstates beyond linear approximation~\cite{Sipe2000,Xiao2010}. 

The complete non-linear conductivity for clean system reads
\begin{align}
    \sigma^{ab;c}&(\bar\omega,\omega_1,\omega_2)
    \notag\\&=
    \frac{-e^3}{\hbar^2 \omega_1\omega_2}
    \sum_{m,n,l}\int_{\bm{k}}
   f_m u_{mm}^{abc}
    \notag\\&
    +f_{mn} \frac{v_{mn}^a w_{nm}^{cb}}{\omega_1+\varepsilon_{mn}}
    +f_{mn} \frac{v_{mn}^b w_{nm}^{ca}}{\omega_2+\varepsilon_{mn}}
    +f_{mn} \frac{w_{mn}^{ab} v_{nm}^c}{\bar\omega+\varepsilon_{mn}}
    \notag\\&
    +\biggl(\frac{f_{mn}  v_{mn}^a v_{nl}^b v_{lm}^c}
    {(\omega_1-\varepsilon_{nm})(\bar\omega-\varepsilon_{lm}
    )}\notag\\ &
    +\frac{f_{mn}  v_{ln}^a v_{nm}^b v_{ml}^c}
    {(\omega_2-\varepsilon_{mn})(\bar\omega-\varepsilon_{ml}
    )}+(a,\omega_1\leftrightarrow b,\omega_2)\biggr),
    \label{eq:generalseco}
\end{align}

where $\bar\omega=\omega_1+\omega_2$.
We point out that Eq.~\eqref{eq:generalseco} differs from  Ref.~\cite{Parker2019} by a sign for the band energy differences in diagrams (III)-(V)~\footnote{private communication}.
As it is visible from the diagrams, the vertices $w$ and $u$ describe instantaneous processes, where two or three interaction events coincide. In contradistinction, diagrams (I) and (II) describe sequential interactions.
We repeat that Eq.~\eqref{eq:sipegeneral} was computed from the dipole Hamiltonian and necessarily assumes the quasiparticle motion to be uninterrupted. Thus it cannot differentiate between instantaneous and successive two-photon processes. 
In the diagrammatics, these processes are easily distinguishable from their vertex structure. 
This distinction can be useful in particular to tell apart effects of the anomalous velocity from the ones produced by higher derivatives. 

\subsection{Finite relaxation rates}
It is possible to prove the equivalence of length and velocity gauge in the adiabatic limit~\cite{Ventura2017}. On a formal level, the equivalence only requires a proper implementation of the time-dependent unitary transformation which translates between state vectors in scalar potential gauge (length gauge) and velocity gauge. 

On the other hand, it is not obvious how to include finite relaxation rates into nonlinear response formulas. 
Dissimilar schemes have been proposed for length and velocity gauge, and in some cases these small changes in the inserted relaxation rates have led to markedly different results for the non-linear response. 
One possibility is to retain different relaxation rates for the non-equilibrium response in each order of the electromagnetic field. Generically, this will produce divergences at resonances due to a mismatch in rate differences~\cite{Cheng2014,Mikhailov2016}.
A careful analysis of this issue was performed recently in Ref.~\cite{Passos2018}. In the reduced density matrix formalism a relaxation rate $\gamma$ is introduced through the appended equation of motion for the density matrix $\rho$, where $\rho_0$ is the equilibrium state,
\begin{align}
    i\hbar \partial_t \rho &=[H,\rho]-i\gamma (\rho-\rho_0).
\end{align}
Translated into velocity gauge this entails a relaxation of the system towards a modified equilibrium distribution $\rho(\bm{A})$~\cite{Ventura2017}. It is not obvious to which physical situation such a relaxation mechanism corresponds to. 
On the other hand, starting from diagrammatics, a commonly employed device is adiabatic switching, which regularizes the poles in the energy denominators by the replacement $\omega\to\omega+i\gamma$. Most importantly, this means inserting $2\gamma$ ($3\gamma,\dots$) in the formulas for second, (third, \dots) harmonic generation whenever $2\omega$ ($3\omega,\dots$) appears. Backporting this prescription to the length gauge, the authors in~\cite{Passos2018} present a case where a highly spiked resonance peak for third harmonic generation becomes noticeably regularized by just this replacement. 

Why is it so difficult to establish a unique procedure?
To demonstrate the difficulties associated with the standard treatment of second order response, 
let us try to interpret the following second order term written in the length gauge ($\varepsilon_{mn}=\varepsilon_m-\varepsilon_n$)~\cite{Sipe2000}
\begin{align}
    \sum_n\partial_{k_c}
    \Bigl(
    \frac{r^a_{mn} v^b_{nm}}{\omega-\varepsilon_{nm}}
    -\frac{v^b_{mn} r^a_{nm}}{\omega-\varepsilon_{mn}}
    \Bigr)
    \label{eq:wrongway}
\end{align}
Taking $\omega=0$, this can be shown to assume the form of the Berry dipole, $\partial^c\Omega^{ab}_{mm}$~\cite{Sodemann2015}. However, performing the momentum derivative first, one obtains instead
\begin{align}
    \sum_n\biggl(&\frac{(\partial_{k_c}r^a_{mn}) v^b_{nm}}{-\varepsilon_{nm}}
    +\frac{r^a_{mn} \partial_{k_c}v^b_{nm}}{-\varepsilon_{nm}}
    +\frac{r^a_{mn} v^c_{nm}v^a_{nm}}{(-\varepsilon_{nm})^2}
    \notag\\&
    -(m \leftrightarrow n)
    \biggr).
\end{align}
Looking at this, it remains unclear which physical process is at play, and the reduction to the Berry dipole by any means other than reverting the derivative is contrived. These issues hold for all such terms, the presence of derivatives in the length gauge formulation make it hard to do power counting in $\omega$ and it is equally laborious to keep track of the proper gauge invariant combinations of terms~\cite{Sipe2000}. The same issues persist for any finite imaginary parts. 

Fundamental to the problem is the resolution of the velocity matrix elements in the eigenbasis, the defining relation which allows to translate response formulas between length and velocity gauge. In the adiabatic limit, one employs $r_{mn}=v^a_{mn}/i\epsilon_{mn}$ ($m\neq n$). The origin of this is the relation
\begin{align}
    v_{mn}^a&=\partial^a\epsilon_{m}\delta_{mn}-i[r^a,H_0]_{mn},
    \label{eq:generalvelocity}
\end{align}
which is invariably true - it follows from the mere fact that the covariant derivative defines a connection in momentum space~\cite{Xiao2010,Parker2019}. However, for a small but nonzero relaxation rate $\Gamma_{mn}$  for excited states, the commutator in this expression is only approximately diagonal in the eigenbasis, and yields
\begin{align}
    v_{mn}^a&=\partial^a\epsilon_{m}\delta_{mn}-i\epsilon_{nm}r^a_{mn}
    \notag\\&\quad
    -\sum_l(r^a_{ml}\Gamma_{ln}-\Gamma_{ml}r^a_{ln})/2.
    \label{eq:velBerryrelation}
\end{align}
Note that $\Gamma_{mn}$ is symmetric. 
The off-diagonal velocity matrix elements are no longer just proportional to the Berry connection. Only in a simplified picture where the relaxation is happening purely between bands $m$ and $n$, i.e. where the relaxation is intraband only, it holds for $m\neq n$ that $r^a_{mn}=v^a_{mn}/i[\epsilon_{mn}+i(\Gamma_{mm}-\Gamma_{nn})]$. 

For the reasons detailed above, we employ the orthodox view of renormalized perturbation theory to approach this issue in the diagrammatic formulation with velocity gauge. Namely, we analyze when the propagators are on-shell or off-shell in the respective diagrams, and use this to introduce  relaxation rates according to the self-energies associated with low-energy or high-energy bands. 
\begin{figure}
\includegraphics[width=.9\columnwidth]{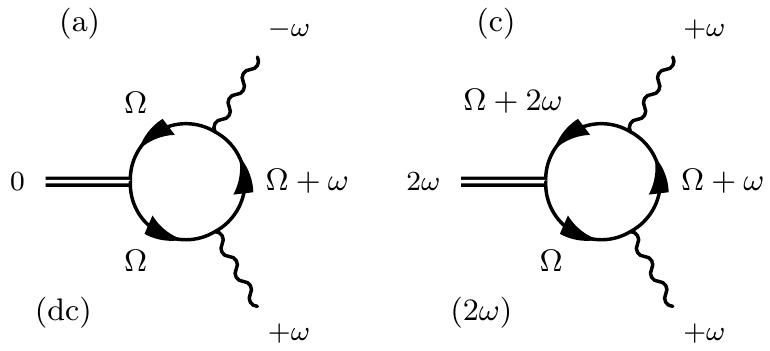}

\includegraphics[width=.9\columnwidth]{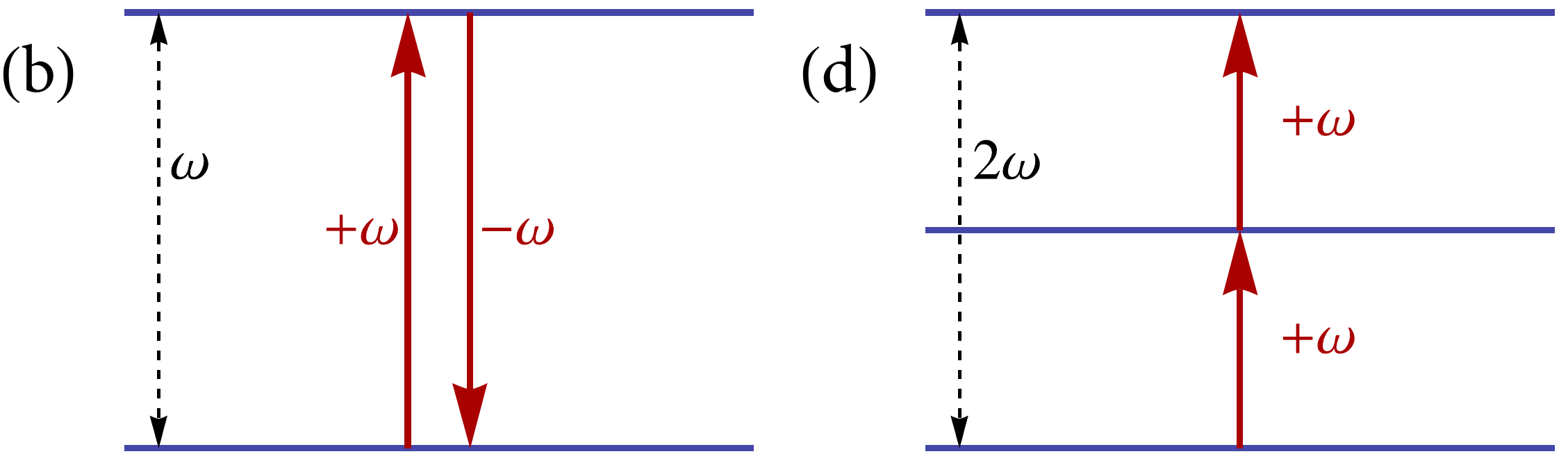}
\caption{Phenomenology of on-shell processes. Diagram (a) for the dc response contains two propagators at the same frequency, while the second harmonic generation shown in (c) involves propagators at three different frequencies. (b) and (d) show these two-photon transitions of the electron in terms of energy levels for either case. In the language of on-shell/off-shell processes, only for diagram (a) (i.e. excitation and de-excitation as indicated in (b)) one encounters a pole which is restricted to intraband relaxation rates near the Fermi surface.}
\label{fig:onshell}
\end{figure}
The approach is physically transparent and does not rely on artificially setting all relaxation rates for all states to be the same. It also sidesteps the issues associated with quasiparticle decay in a density matrix formulation, instead relying on the Green's function formalism to implement finite lifetimes.
More importantly, different relaxation rates appear depending on the type of process and not on the order in perturbation theory. This also means that second harmonic generation does not involve the same relaxation rates as dc-current generation.
For example, the three-point loop  depicted in Fig~\ref{fig:onshell} involves two consecutive photon processes, which means the quasiparticle partially propagates  in highly excited bands. However, comparing the on-shell processes ($\Omega=0$), for the photovoltaic effect ($\omega_1=-\omega_2$) the quasiparticle propagators are dissimilar to the case of second harmonic generation ($\omega_1=\omega_2$). It  seems consequent to take this into account by using appropriate (i.e. dissimilar) relaxation rates. In the calculation, this reasoning calls for the insertion of a constant self-energy representative of the respective pole location where the residue is taken. We stress that such a procedure is insensitive to the analytic continuation to real frequencies as the actual self-energy is still only a constant, just not the same constant in all propagators. Further details of the relaxation time approximation are summarized in App.~\ref{app:relax}.

In short, the presented phenomenology can be seen as the replacement $\omega\to\omega+i\gamma$, with the additional stipulation that the relaxation rate $\Gamma$ of the intermediate states is allowed to be different from relaxation rate $\gamma\ll \Gamma$ of the long-lived quasiparticles close to the Fermi surface. We emphasize that it is certainly possible to systematically include the very same phenomenology in the density matrix approach with length gauge, but it seems to lack the intuitive interpretation available in the diagrammatic analysis.

While we do not specify the microscopic nature of the relaxation rates, it is important to point out that $\gamma$ is expected to be given by the transport time, and not the quasiparticle lifetime. This is related to the fact that whenever $\gamma$ becomes important, the relevant processes happen at the Fermi surface where the usual arguments apply for the velocity vertex renormalization~\cite{Lee1985}. In contrast, $\Gamma$ is an interband quantity which is unrelated to the scattering processes happening at the Fermi surface.

\section{Anomalous acceleration}
In order to intuitively understand second order response in the absence of $\mathcal{T}$, one might want to fall back on a semiclassical treatment of the problem.

Disregarding lifetime effects and assuming $\mathcal{T}$, the semiclassical result for the non-linear conductivity is stated as~\cite{Morimoto2016a,Parker2019}
\begin{align}
\sigma^{ab;c}&=\frac{e^3}{\hbar^2}\int_{\bm{k}}
\sum_m f_{m}\left(
-i\frac{\partial^b\Omega^{ca}_{mm}}{\omega}
\right),
\label{eq:gappedSC}
\end{align}
where the relaxation rate can be reintroduced after the fact by the replacement $\omega\to \omega+i\gamma$. The Abelian Berry curvature is defined as $\Omega^{ab}_{mm}=
\partial^ar_{mm}^b-\partial^br_{mm}^a
$.

However, this phenomenology breaks down once the result from perturbation theory is considered. Firstly, the usual semiclassical derivation cannot account for processes which are instrinsically finite-$\omega$, but non-negligible after the momentum space integral. Both the shift current and the injection current are of this type. Secondly, non-linear response can involve interband transitions which are easily missed in the construction of the collision integral.
The situation is complicated by the fact that finite lifetimes also modify the relation between Berry connection and velocity matrix elements according to Eq.~\eqref{eq:velBerryrelation}, as we discussed before. 

\subsection{Vertices and sum rules}

The calculation proceeds as follows. 
We first note that the derivative of the Berry connection is $\partial^b r^c_{mn}=R^{bc}_{mn}+i[r^b,r^c]_{mn}$, where $2R^{bc}_{mn}=\langle i \partial^b\partial^c m| n\rangle+\langle m| i \partial^b\partial^c n\rangle$, as mentioned. This matrix element can be related to the quantum metric through $2g^{ab}_{mn}=2iR^{ab}_{mn}
+2\langle m| \partial^a\partial^b n\rangle
	-\sum_l f_l (r^a_{ml}r^b_{ln}
	+ r^b_{ml}r^a_{ln})$.
The Berry dipole is given by [cf.~Sec.~\ref{app:A}]
\begin{align}
	\partial^ a\Omega^{bc}_{mm}
	&=-\tfrac{1}{2}[r^a, [r^b,r^c]]
	+i[r^b,R^{ac}] - i[r^c,R^{ab}],
\end{align}
where all matrix elements carry two band indices which are circled through as dictated by the commutators (i.e. $[a,b]_{mn}=\sum_{l}(a_{ml}b_{ln}-b_{ml}a_{ln})$). We will henceforth suppress band indices whenever possible. 
Furthermore for interband terms it holds that $i\partial^cv^b=-\varepsilon\partial^cr^b-\Delta^cv^b+\mathcal{O}(\Gamma)$. These relations, while not exact for finite relaxation rates, are perturbative in relaxation rate corrections and do not introduce any shifted poles.

Sometimes, it is suggested to use the substitution $r^a_{mn}\to i v^a_{mn}/\varepsilon_{nm}$. However, as explained earlier, such is only possible in a clean system.
This is noteworthy since the decomposition of $w^{ab}_{mn}$ creates, among others, terms which are very similar to diagrams (I) and (II).
Going forward, we will only employ the opposite route of identification via $v^a_{mn}\to i \varepsilon_{mn}r^a_{mn}+\mathcal{O}(\Gamma)$, which always remains viable. This allows to immediately and transparently characterize the role of sum rules for systems with finite lifetimes and to establish a unique procedure to translate between the different gauges.

We rewrite $w^{ab}$ and $u^{abc}$ by expanding the derivatives,
\begin{align}
    w^{ab}_{mn}&=\partial^av^b-i[r^a,v^b]
    \qquad \text{for } m\neq n\notag\\
    &=\tfrac{1}{2}i([v^a,r^b]
    +\Delta^a r^b+\varepsilon R^{ab}+a\leftrightarrow b)+\mathcal{O}(\Gamma)
    \label{eq:sumrule1}
    \\
    u^{abc}_{mm}&=
    \partial^a(\partial^bv^c-i[ r^b,v^c])
    -i[ r^a,\partial^b v^c-i[r^b,v^c]]
    \notag\\
    &=\partial^a\partial^bv^c
    +i[v^c,R^{ab}]+i[v^b,R^{ca}]+i[v^a,R^{bc}]
    \notag\\&\quad
    +[r^b,\Delta^a r^c]+[r^a,\Delta^b r^c]
    -[r^a,[r^b,v^c]]
    +\mathcal{O}(\Gamma).
    \label{eq:sumrule2}
\end{align}
The decomposition of the vertices might still appear pretty verbose, but crucially it will allow for a straightforward identification of the physical processes. 
Independent of this, it makes manifest that relaxation rates are accounted for perturbatively in the velocity expansion of the higher order vertices.
As an aside, note that only the diagonal component of $u^{abc}$ was analyzed since this is the only manner in which it appears here. The off-diagonal elements of $u^{abc}$ needed for higher orders  follows straightforwardly by continuing the expansion of $\partial^a\partial^bv^c$ for off-diagonal matrix elements.
Keeping the shorthand notation, the second-order response is generally
 \begin{align}
    \sigma^{ab;c}&=\frac{-e^3}{\hbar^2\omega_1\omega_2} \int_{\bm{k}}\sum_m
    f_{m}\biggl(
    u^{abc}+[w^{ab},\frac{v^c}{\bar\omega-\varepsilon+i\gamma}]
    \notag\\&\quad
    +[\frac{v^a}{\omega_1+\varepsilon+i\Gamma},w^{bc}]
    +[\frac{v^b}{\omega_2+\varepsilon+i\Gamma},w^{ca}]
    \notag\\&\quad
    +[[v^b,\frac{v^c}{\varepsilon-\bar\omega-i\gamma}],\frac{v^a}{\omega_1+\varepsilon+i\Gamma}]
    \notag\\&\quad
    +[[v^a,\frac{v^c}{\varepsilon-\bar\omega-i\gamma}],\frac{v^b}{\omega_2+\varepsilon+i\Gamma}]\biggr)_{mm}.
    \label{eq:generalsemics}
\end{align}
While all top level commutators carry the indices $mm$, they do not evaluate to zero since the Fermi-Dirac factor $f_m$ is not incorporated, i.e. the expression is not a trace.
Nevertheless, thanks to the Fermi factor any top level commutators actually evaluate to zero for the diagonal pieces where a denominator $\varepsilon=0$ could otherwise lead to issues. This allows the replacement of velocity matrix elements by the Berry connection in outer commutators and in particular renders the $i\gamma$ in the first line harmless. In principle, the frequencies $\omega_1$ and $\omega_2$ in front of the integral also acquire an imaginary part according to adiabatic switching ($\omega\to\omega+i\gamma$), but this replacement does not lead to any further complications and will henceforth be implicitly assumed for clarity of presentation.

The usual simplifications for a gapped and time-reversal symmetric system resulting in Eq.~\eqref{eq:gappedSC} are implemented using the sum rules Eqs.~(\ref{eq:sumrule1},\ref{eq:sumrule2}), which present the proper extensions of the classical sum rules for non-quadratic Hamiltonians in systems with finite quasiparticle lifetimes~\cite{Aversa1995,Sipe2000,Ventura2017}.
$\mathcal{T}$ dictates that velocity matrix elements fulfill $v^a_{mn}(\bm{k})=-v^a_{mn}(-\bm{k})^*$, while the non-Abelian Berry connection transforms like $r^a_{mn}(\bm{k})=r^a_{mn}(-\bm{k})^*$, and $R^{ab}_{mn}(\bm{k})=-R^{ab}_{mn}(-\bm{k})^*$. For a gapped system the dc conductivity follows from the simultaneous expansion in small $\bar\omega\ll\omega_1,\omega_2$; we will also keep track of finite lifetimes whenever necessary. 
The integrand of the leading order term of size $\omega^{-2}$ is explicitly ($\delta^c_{mn}=r_{mm}-r_{nn}$)
\begin{align}
    &u^{abc}-i[w^{ab},r^c]+i[r^a,w^{bc}]+i[r^b,w^{ca}]
    \notag\\
    &-[[v^a,r^c],r^b]-[[v^b,r^c],r^a]
    -[v^a\delta^c,r^b]-[v^b\delta^c,r^a]
    \notag\\
    &+\tfrac{1}{\gamma}[v^a\Delta^c,r^b]+\tfrac{1}{\gamma}[v^b\Delta^c,r^a]
    \label{eq:wsquaredfull}
\end{align}
Here, we used that $[v^a,v^c/(\varepsilon-i\gamma)]\approx i[v^a,r^c]+iv^a\delta^c-iv^a\Delta^c/\gamma$.
We reiterate that sum rules basically constitute expansions of the vertices $w$ and $u$ in terms of acceleration ($w$) and jerk ($u$). Accordingly, the sum rules contain commutators with exactly the required number of odd components under momentum inversion. For example, it is
\begin{align}
    [v^c(-\bm{k}),R^{ab}(-\bm{k})]
    &=(-1)^2[v^c(\bm{k})^*,R^{ab}(\bm{k})^*]\\
    &=-[v^c(\bm{k}),R^{ab}(\bm{k})]\\
    [r^a(-\bm{k}),\Delta^c(-\bm{k})r^b(-\bm{k})]
    &=-[r^a(\bm{k})^*,\Delta^c(\bm{k})r^b(\bm{k})^*]
    \notag\\
    &=-[r^a(\bm{k}),\Delta^c(\bm{k})r^b(\bm{k})].
\end{align}
Unlike in linear response, the Drude-type term $\partial^a\partial^bv^c$ is odd under momentum inversion and equally cancels.
In essence, in the presence of $\mathcal{T}$, higher order vertices do not contribute at leading order in $\omega$ to $\sigma^{ab;c}$. However, from the definition of the vertices ($w^{ab}_{mn}=\langle m| \partial^a\partial^b H|n\rangle$ and $u^{abc}_{mm}=\langle m| \partial^a\partial^b\partial^c H|m\rangle$), the same conclusion can be drawn immediately from their symmetry properties under momentum inversion, thus sidestepping the need  for sum rules.

\subsection{Finite frequency terms}
The last two lines in Eq.~\eqref{eq:generalsemics} require more care as they contain a relaxation rate in the nested denominator, where a finite $\gamma$ is not zeroed out a priori. While Eq.~\eqref{eq:wsquaredfull} suggests that both terms proportional to $\gamma^{-1}$ cancel each other, this is not necessarily the case as the limit $\omega\to 0$ does neither commute with the k-space integration nor with the limit $\Gamma\to 0$. Thus, while Eq.~\eqref{eq:wsquaredfull} correctly captures the response at $\omega=0$, it misses not only the injection current, but also the shift current, both are by construction finite frequency responses. Their incorporation is of course straightforward by keeping all imaginary parts in Eq.~\eqref{eq:generalsemics}. Denominators containing $\Gamma$ have a non-zero imaginary part for $\varepsilon=\omega_{1,2}$ and lead to the shift current, while denominators with $\gamma$ lead to the injection current.
\citeauthor{Sipe2000}~\cite{Sipe2000} discussed the order of limits $\gamma\to 0$, and then $\bar\omega\to 0$. The divergence then manifests in a finite current injection, $\dd J/\dd t\neq 0$. Another alternative is to discuss the dc case as the limit $\bar\omega\to 0$, followed by $\gamma\to 0$~\cite{Parker2019}.
In this context it is important to keep in mind that any finite quasiparticle lifetime will limit the phenomenon of current injection to  irradiation times shorter than the lifetime. At longer times, the current instead saturates at a large but finite current. For simplicity, we continue to call a current proportional to $\gamma^{-1}$ \emph{injection} current, even when keeping $\gamma$ finite.

The injection current is due to the conductivity
\begin{align}
\sigma^{ab;c}_{(i)}&=\frac{e^3}{\hbar^2\omega^2}\int_{\bm{k}}
\sum_m f_{m}
\frac{[v^a\Delta^c, v^b \mathcal{D}]_{mm}}{i\gamma}
\label{eq:correctinj}
\\
\mathcal{D}_{mn}&=\frac{1}{\varepsilon_{mn}+\omega+i\Gamma}+\frac{1}{-\varepsilon_{mn}-\omega+i\Gamma},
\end{align}
In the clean limit the last term simplifies to the injection current in the form written in Eq.~\eqref{eq:sipegeneral}. In particular, for applied frequencies $\omega\gg \Gamma$, $\mathcal{D}$ assumes the form of $\delta$-functions. This means that current injection will occur invariably for any ordering of the three limits $\bar\omega\to0$, $\gamma\to0$ and $\omega\to0$.
However, completing the expansion under the assumption that $\omega\ll \Gamma$, the result is instead
\begin{align}
\sigma^{ab;c}_{(i)}
&=\frac{e^3}{\hbar^2\omega^2}\int_{\bm{k}}
\sum_m f_{m}
[r^a\Delta^c, r^b ]_{mm}
\frac{-2\Gamma}{\gamma}.
\end{align}
This expression is finite as long as  $\Gamma/\gamma<\infty$, meaning that the effect of current injection ceases in the strict zero frequency limit. 
We caution that the parameter regime $\omega\ll \Gamma$ is not well controlled in the present formulation, which somewhat hides the effects of Pauli blocking and disregards possible localization effects, both of which might further decrease the current.
However, this does not affect our observation that it requires not only time-reversal breaking, but also well defined quasiparticles ($\omega\gg \Gamma$) to prevent the cancellation of the time-reversed paths. 
The effect of current injection for ($\omega\gg \Gamma$) is reminiscent of antilocalization in a disordered system with spin-orbit coupling. 
In systems with antilocalization, pairs of counterpropagating paths will not interfere destructively due to the spin-momentum locking, leading to an increased value of the linear conductivity. Upon removing spin-orbit coupling, both paths acquire opposite Aharonov-Bohm phases and the destructive interference between them leads to a decrease in conductivity. 
Notably, weak antilocalization persists in the strict adiabatic limit.
In contrast, the effect of current injection discussed here relies on applied frequencies which exceed the inverse lifetime of excited states ($\omega\gg\Gamma$), and is thus intrinsically a dynamical effect, which does not extend to the adiabatic limit where all frequencies approach zero uniformly. 

\begin{figure}
\includegraphics[width=.7\columnwidth]{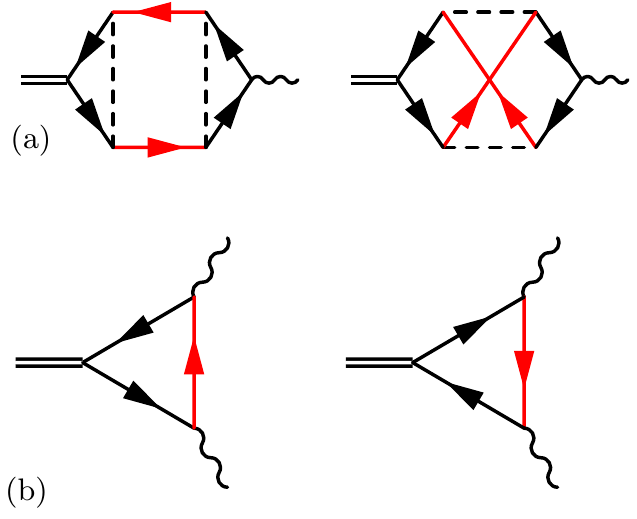}
    \caption{Comparison of the lowest order diagrams which contain time-reversed paths for the case of (a) impurity scattering and (b) electromagnetic interaction. Dashed lines denote disorder lines, the red colored propagators are coherently counter-propagating. The disorder problem does not have any lower order diagrams of this kind because the disorder average removes any correlations between self-energy terms of different propagators, making it necessary for the disorder to connect two fermion lines in order for it to have a coherent effect.}
    \label{fig:weakanti}
\end{figure}

The common property which we want to draw attention to is the following: Both in weak localization and for the dynamical effect discussed here the leading order contribution to the conductivity is (fully or partially) neutralized due to destructive interference with another term where a retarded propagator is exchanged for an advanced one, \emph{unless} the phase of the electron wavefunction is rotated non-trivially (for example due to time-reversal breaking). 
Crucially, the lowest order where this happens for weak antilocalization is the crossed disorder diagram, but for second order optical response it is the three-leg loop (Fig.~\ref{fig:weakanti}). The effect first appears in different diagrams due to the fact that self-energy corrections from disorder interactions are not coherent between time-reversed paths, whereas the interaction with an external plane wave is coherent. Another difference is that the treatment of localization effects require the summation of the Diffuson and Cooperon~\cite{Lee1985}, while the finite-frequency transitions considered here are perturbative in the applied electric field.

Current injection could therefore be described as an effect of \emph{dynamical} antilocalization. It is tied to the appearance of fermion loops with more than two vertices, meaning similar effects should be present in non-linear response functions of any order, given some suitable order of limits for the external frequencies. 

Using the expression for the clean case, the $\delta$-functions allow the substitution $\omega\to-\varepsilon_{mn}$ in the global coefficient. The result coincides with the corresponding expression in Eq.~\eqref{eq:sipegeneral},
\begin{align}
\sigma^{ab;c}_{(i)}&=\frac{-e^3}{\hbar^2}\int_{\bm{k}}
\sum_m f_{m}
\frac{[r^a\Delta^c, r^b \mathcal{D}]_{mm}}{i\gamma}.
\end{align}

Then the following suggestive formulation of the commutator is possible:
\begin{align}
    [r^a\Delta^c, r^b\mathcal{D}]&=[v^a,\partial^cr^b\mathcal{D}]-[r^a,\partial^cv^b\mathcal{D}]
\end{align}
where we used  $\partial^cv^b=i\varepsilon\partial^cr^b+i\Delta^c r^b$. 
We therefore interpret the injection current as  a mismatch between $v^a\partial^cr^b$ which contains the derivative of the Berry connection $\partial^cr^b$, and $r^a\partial^cv^b$ due to the kinematic acceleration $\partial^cv^b$, each of them integrated over the manifold where $\varepsilon_{mn}=\pm \omega$. 
This reinforces the dynamical nature of the injection current, which appears due to the different accelerations acting on a band wavefunction and on a quasiparticle.
It is not possible to formulate either acceleration purely in terms of the Berry curvature, with the exception of a two band model, where $\Delta^c_{mn}$ can trivially be moved outside of the commutator~\cite{deJuan2017}.

The same treatment can be applied to the shift current by analyzing resonances for $\Gamma\to 0$, which yields
\begin{align}
    \sigma^{ab;c}_{(s)}
    &=\frac{e^3}{\hbar^2\omega^2}\int_{\bm{k}}
    \sum_m f_{m}
    \bigl(
    [i\pi\delta(\omega+\varepsilon)v^a,\tfrac{\varepsilon}{2}[r^b,r^c]
    \notag\\&\quad
    -iv^b\delta^c+i\varepsilon R^{bc}]
    +[i\pi\delta(\omega+\varepsilon)v^b,\tfrac{\varepsilon}{2}[r^a,r^c]
    \notag\\&\quad
    -iv^a\delta^c+i\varepsilon R^{ac}]
    \bigr)_{mm}.
\end{align}
This expression for the shift current is identical to the established formulation in terms of a shift vector~\cite{vonBaltz1981}. For linear polarized light it is once again helpful to insert the band energy differences (i.e. replace $\omega\to\pm\varepsilon_{mn}$), to obtain for finite relaxation rates
	\begin{align}
	    \sigma^{aa;c}_{(s)}
	    &=\frac{e^3}{\hbar^2}\int_{\bm{k}}
	    \sum_m f_{m}
	    \bigl(
	    -[\mathcal{D}^{'}r^a,R^{ca}]
	    \notag\\&\quad
	    +i[\mathcal{D}^{'}r^a,\tfrac{1}{2}[r^a,r^c]+\delta^cr^a]
	    \bigr)_{mm}.
	    \label{eq:correctshift}
	    \shortintertext{where we used}
	    \mathcal{D}_{mn}^{'}&=\frac{-1}{\varepsilon_{mn}+\omega+i\Gamma}+\frac{-1}{\varepsilon_{mn}-\omega+i\Gamma},
	\end{align}
The shift current can thus be viewed as a geometric effect with two sources, the quantum metric and the motion due to the Berry dipole, but with the intraband acceleration projected out. For an explanation of this second part, we recall that $\delta^cr^a$ appears as a consequence of replacing the velocity $v^c$ in Eq.~\eqref{eq:generalsemics} by the Berry connection, compensating the overcounting that happens with respect to the intraband motion. This intraband motion is actually diabatic and is exactly what gives rise to the injection current; consequently the projected quantity $[r^c,r^a]+\delta^cr^a$ comprises only the remaining adiabatic processes. 
We emphasize that the obtained expression for the shift current is in agreement with the previous literature~(e.g.~\cite{Sipe2000}). At the same time, the presented derivation of a counterterm for the diabatic motion which forms part of the shift current allows us to connect this term with the ballistic current from the kinetic approach: Just like the ballistic current, it carries a factor $1/\gamma$ from the injection current, which is cancelled by the expansion in $\gamma/\varepsilon$~\cite{Deyo2009,Sturman2019}. 
This confirms that the theory of second-order optical response recovers such an order-one effect in the shift current, an important consistency check. 
We point out that in the literature, the shift current is sometimes further subdivided into ballistic current $\bm{j}_b$, shift current due to excitation $\bm{j}_{ex}$, and the shift current due to recombination $\bm{j}_{rec}$~\cite{Sturman2019}. This is necessary if the source of the shift is not intrinsic, i.e. not due to the asymmetry in the band structure but due to impurity scattering. In this case, the lifetimes entering Eq.~(35) can be dissimilar so that the sum of these three currents is no longer uniquely determined by the dispersion. The same observation was made by considering the side jump mechanism as a source of electron shift~\cite{Deyo2009}.
As a study of microscopics is beyond the scope of our work, it is not possible to check whether all results of the kinetic approach are recovered in the shift current formalism for finite lifetimes. This is left for future work.

In summary, borrowing from the language of antilocalization, current injection occurs because there is a mismatch between adiabatic motion and finite frequency motion, which by the order of limits which defines optical response cannot cancel. We repeat that $\mathcal{T}$ enforces that current injection appears only in response to circular polarized light. The shift current is in turn the sum of all remaining finite frequency processes, except for the type of intraband motion which already gives rise to the injection current. 
The fundamental difference between shift and injection current can also be discussed in terms of the units of the involved integrands. While the matrix elements in the shift current, Eq.~\eqref{eq:correctshift} naturally form a quantity with the units of a Berry dipole, this is not true for the injection current, because there is an additional relaxation time in the denominator. This distinction will also prove useful in the semiclassical expressions.

\subsection{Anomalous semiclassical terms}
Finally, we derive the semiclassical expansion in the absence of $\mathcal{T}$. To repeat, in this case  Eqs.~(\ref{eq:sumrule1},\ref{eq:sumrule2}) are no longer helpful in canceling even powers in the $\omega$-dependence. Up to order $\omega^{-1}$, the regular terms are
\begin{align}
    \sigma^{ab;c}&=\frac{e^3}{\hbar^2}
    \int_{\bm{k}}\sum_m
    f_m\biggl(
    \frac{S_{-2}}{\omega^2}+\frac{S_{-1}}{\omega}
    +\dots\biggr)
    \label{eq:fullsemiclassical}
    \\
    S_{-2}&=\partial^a\partial^b v^c
    -[r^a,\Delta^c r^b]
    -[r^c,[v^a,r^b]]
    +[r^c,[r^a,v^b]]
    \\
    S_{-1}&= -i\partial^c\Omega^{ab}-2[\tfrac{r^a}{\varepsilon},\Delta^c r^b]
\end{align}
As explained, injection and shift current do not appear here as they are not amenable to an expansion in small $\omega$. Also, the leading correction for finite lifetimes is given by the replacement $\omega\to\omega+i\gamma$.
From the systematics of expression Eq.~\eqref{eq:generalsemics}, one can immediately deduce that only the term of order $\omega^{-1}$ can have a purely geometric origin. All other orders necessarily depend on velocity matrix elements or the dispersion relation. Importantly, for $\mathcal{T}$ we recover immediately that $S_{-2}=0$, and $S_{-1}$ becomes an only function of the Berry dipole $\partial^c\Omega^{ab}$. 
This latter term is the well-known non-linear AHE~\cite{Deyo2009,Sodemann2015,Morimoto2016a}. 
We emphasize that in the absence of $\mathcal{T}$, there is also an AHE at linear order in the electric field. In this case, it can be difficult to distinguish linear from non-linear contributions. At least for the intrinsic contributions to the AHE discussed here, the leading order non-linear effect as given by $S_{-2}$ scales with $\gamma^{-2}$, which is distinct from the linear AHE, which is of order one. Alternatively, in materials which break both inversion symmetry ($\mathcal{P}$) and $\mathcal{T}$ but preserve the combined symmetry $\mathcal{PT}$, the Berry curvature is identically zero everywhere, which means that the linear AHE vanishes.

The second order response, Eq.~\eqref{eq:generalsemics} thus decomposes into two pieces given by Eq.~(\ref{eq:correctinj},\ref{eq:correctshift}) with a singular frequency dependence and the regular terms, Eq.~\eqref{eq:fullsemiclassical}. Compared to previous results, the reported formulas have no further derivatives which act on the propagators, they explicate the frequency dependence and they phrase the response in terms of four physical processes, each of which is, in principle, measurable independently. 
For the limit $\omega\to 0$, our semiclassical expressions recover the lifetime dependence which was found for the non-linear anomalous Hall conductivity as calculated in the Boltzmann kinetic approach~\cite{Deyo2009,Gao2019}.
As an aside, we point out that the most often used formula for shift and injection current, Eq.~\eqref{eq:sipegeneral}, does not lend itself to a straightforward generalization for finite lifetimes, as the replacement of applied frequencies by band energy differences is only possible in the limit $\omega\gg \gamma$. Instead, the correct starting point in the length gauge are expressions containing the covariant derivative, which can be found for example in~\cite{Ventura2017,Passos2018}.

In the absence of a gap, the semiclassical expansion, Eq.~\eqref{eq:fullsemiclassical} for small $\omega$ becomes questionable. This is, however, not a problem for the wide range of Dirac and Weyl semimetals which have undertilted cones: The regions in k-space in the direct vicinity of the Dirac points do not contribute significantly to the non-linear conductivity, rendering them irrelevant. In contrast, in type II Weyl semimetals this does not hold, hence they exhibit a very different phenomenology~\cite{Yang2017}.

\subsection{Conductivity tensor}
At this point it is helpful to recall that sometimes several components of the non-linear conductivity $\sigma^{ab;c}$ are combined for a given current response, depending on the applied electric fields. For example, the BPVE for linear polarized light with polarization direction along the $x,y$-diagonal is given by
\begin{align}
    j^c&=\tfrac{1}{2}(\sigma^{xx;c}+\sigma^{yy;c}+\sigma^{xy;c}+\sigma^{yx;c})|\bm{E}|^2.
\end{align}
In the same fashion, light with anticlockwise polarization in the $x,y$-plane leads to the current
\begin{align}
    j^c&=\tfrac{1}{2}(\sigma^{xx;c}-\sigma^{yy;c}+\sigma^{xy;c}-\sigma^{yx;c})|\bm{E}|^2.
\end{align}
We will employ some of these properties in the next section to recast the discussion in a more familiar fashion in terms of the symmetries of real and imaginary parts of the matrix elements.

\section{dc current response}
In the following we explore the dc-current creation from a single Weyl cone and relate it to the general phenomenology we explained in the previous section. 
Of particular interest will of course be the injection current, but we also make some comments regarding the remaining contributions which combine to the shift current.
Diagrams (III)-(VI) contain the contribution from curvature effects in the intraband motion, they do not contribute to the injection current and will therefore be disregarded in this section.

Since the Abelian Berry curvature is strictly zero in a space-time inversion (combined symmetry by the inversion symmetry and $\mathcal{T}$, $\mathcal{PT}$) symmetric material, the contribution from the Berry dipole vanishes. We utilize this special case to investigate the dc-response when it is dominated by acceleration terms unrelated to the Berry curvature.

To repeat, the second-order optical response is given by the six diagrams depicted in Fig.~\ref{fig:diags}. The dc-current response is  $\sigma^{ab;c}(0,\omega,-\omega)\equiv \sigma^{ab;c}_{dc}$.
Additionally, for the benefit of familiarity, in this section we will fall back on the standard notation in band indices instead of the very compact notation of the previous section.
Without loss of generality, consider in the following a coordinate system which is aligned with the direction of the incident linear polarized light. Then  for the circular polarization, the indices $a$ and $b$ of the electric field are dissimilar while for the linear polarized light it is $a=b$. For any other choice of coordinates, the following statements still hold for the respective current response, but not individually for the matrix elements.
It follows in either case that
\begin{align}
    \sigma^{ab;c}_{circ}
    &=\sum_{m,n,l}\int_{\bm{k}}f_{nl}\,\mathrm{Im}\left[A^{abc}_{nml}(\omega)-A^{abc}_{nml}(-\omega)\right]\\
    \sigma^{aa;c}_{lin}
    &=\sum_{m,n,l}\int_{\bm{k}}f_{nl}\,\mathrm{Re}\left[A^{aac}_{nml}(\omega)+A^{aac}_{nml}(-\omega)\right].
\end{align}
Furthermore,
\begin{align}
A^{abc}_{nml}(\Omega)= & \frac{ e^3}{\hbar^2\Omega^2}\frac{N_{nml}^{abc}}{(\varepsilon_{mn}-i\gamma)(\Omega-\varepsilon_{nl}-i\Gamma)}
\end{align}
where $N_{nml}^{abc}=v_{nl}^a v_{lm}^b v_{nm}^c$. Note that the intraband relaxation rate $\gamma$ is usually  significantly smaller than the interband relaxation rate $\Gamma$. 
The formulation as real and imaginary part follows form the observation that the incoming electric fields either fulfill $E^a(\omega)E^b(-\omega)=E^a(-\omega)E^b(\omega)$ (linear polarization) or $E^a(\omega)E^b(-\omega)=-E^a(-\omega)E^b(\omega)$ (circular polarization).
We will occasionally use the shorthand $\Sigma^{ab;c}_{\pm}(\omega)$ with $\sigma^{abc}_{lin}=\Re\Sigma^{ab;c}_{+}(\omega)$ and $\sigma^{abc}_{circ}=\Im\Sigma^{ab;c}_{-}(\omega)$

\subsection{Cancellations from time reversal symmetry}
\label{sec:circtrs}
In the presence of $\mathcal{T}$, energy eigenvalues are inversion symmetric in the Brillouin zone, $\varepsilon_n(\bm{k})=\varepsilon_n(-\bm{k})$ and velocity matrix elements fulfill $v_{nm}^a(\bm{k})=-{v_{nm}^a}^*(-\bm{k})$. In this case, both injection current and shift current benefit from cancellations between $\bm{k}$ and $-\bm{k}$. The important term to keep track off is $\varepsilon_{mn}-i\gamma$, which in the limit $\gamma\rightarrow 0$ is manifestly divergent for $m=n$.
We therefore proceed to write out the real and imaginary parts of the energy denominators $D_{mn}^{-1}=\varepsilon_{mn}-i\gamma$ and $D_{nl}^{-1}(\omega)=\Omega-\varepsilon_{nl}-i\Gamma$. Since the nonlinear dc-response is proportional to $E_a(-\omega) E_b(\omega)$ one can replace $a\leftrightarrow b$ and $\omega\leftrightarrow-\omega$ in the expression for $A$, meaning that for an inversion symmetric Brillouin zone the second order response can be written as proportional to
\begin{align}
N_{nml}^{abc}- s N_{mnl}^{abc}&=V_s,
\end{align} 
where $s=+1$ ($s=-1$) for linear (circular) polarized light. 
For the various combination possibilities of real and imaginary parts of $J^{lin}$, this implies
\begin{align}
    \Re D_{mn} \Re D_{nl}(\omega) \Re V_1 &=0\label{Eq:lin1}\\
    -\Im D_{mn} \Re D_{nl}(\omega) \Im V_1     &=0\label{Eq:lin2}\\
    -\Re D_{mn} \Im D_{nl}(\omega) \Im V_1     &\neq 0,\label{Eq:shiftTRS}\\
    -\Im D_{mn} \Im D_{nl}(\omega) \Re V_1 &=0\label{Eq:lin4}
\end{align}
while $J^{circ}$ has only one cancellation,
\begin{align}
    \Re D_{mn} \Re D_{nl}(\omega) \Im V_{-1} &\neq 0\label{Eq:circ1}\\
    \Im D_{mn} \Re D_{nl}(\omega) \Re V_{-1}     &= 0\label{Eq:injectionTRS}\\
    \Re D_{mn} \Im D_{nl}(\omega) \Re V_{-1}     &\neq 0.\label{Eq:injectionTRS2}\\
    -\Im D_{mn} \Im D_{nl}(\omega) \Im V_{-1}  &\neq 0.\label{Eq:circ4}
\end{align}
In Eqs.~(\ref{Eq:lin1},\ref{Eq:lin4}) we made use of the fact that $a=b$ for linear polarized light, which implies $\Re V_1=0$. 
Eq.~(\ref{Eq:lin2}) follows from the fact that $\Im D_{mn}\neq 0$ requires $m=n$, which entails $\Im V_{1}=0$, and Eq.~\eqref{Eq:injectionTRS} is zero due to the antisymmetry in the band summation for $m=n$.

\subsection{Space-time inversion symmetry}
In the absence of $\mathcal{T}$, the aforementioned cancellations do no longer hold and a comparable behavior (injection current) is expected for both linear and circular polarization. Using the conventional argumentation, without $\mathcal{T}$ any combination of real and imaginary parts of the denominators is potentially nonzero and has to be reconsidered. This is of course nonwithstanding crystal symmetries, which can still render certain components zero.
As we discussed, the well-known formula for the shift current and linear polarization, 
\begin{align}
    \sigma^{aa;c}_{lin}&\sim
    \sum_{n,m} f_{nm}\Im[r_{mn}^ar_{nm;c}^a]\delta(\varepsilon_{nm}-\omega),
\end{align}
is no longer the most relevant part of the response function once $\mathcal{T}$ is broken. This term originates from Eq.~\eqref{Eq:shiftTRS} only and evaluates to zero for $m=n$. Instead, resonances involving $m=n$ provide the dominant contribution to the BPVE. 
It is then useful to separate two-band contributions containing $\Im D_{mn}=\gamma^{-1}\delta_{mn}$ from three-band ones with $\Re D_{mn}=\varepsilon_{mn}^{-1}(1-\delta_{nm})$, the latter of which are suppressed by the energy denominator $\varepsilon_{mn}$. 

In the presence of $\mathcal{PT}$-symmetry, only the real part of the product of momentum matrix elements $N_{nml}^{abc}$ contributes.
Imposing $\mathcal{PT}$ and focusing on the dominant two-band contributions, the response therefore features the following term 
\begin{align}
    -\sum_{m,n,l}f_{nl}\Im D_{mn} \Im D_{nl}(\omega) N_{nml}^{aac}\pm(\omega\rightarrow-\omega).
\end{align}
Note that for a $\mathcal{T}$ material, only the imaginary part of the numerator $N_{nml}$ contributes to the dc-current, while in the presence of $\mathcal{PT}$-symmetry only the real part does. In this sense, both are  complementary.
Since $\mathcal{T}$ is broken, a symmetrization with respect to $\bm{k}$ is no longer useful.
Instead, one should consider each Dirac cone separately. It was previously noted that semimetals have zero shift current if the cone has a linear dispersion~\cite{Yang2017,Chan2017}. This is related to the triple product of velocities, which is antisymmetric on a radial shell which concentrically encloses a perfect Dirac point. 

To be concrete, suppose a tight binding model with $\mathcal{PT}$-symmetry, constituted of two pairs of degenerate bands. Given a trivial implementation of inversion symmetry, the most general $\mathcal{PT}$-symmetric four-band Hamiltonian is~\cite{Tang2016}
\begin{align}
    H_{PT}=
    g_{x0}(\bm{k})\tau_x
    +g_{z0}(\bm{k})\tau_z
    +g_{yx}(\bm{k})\tau_y\sigma_x\notag\\
    +g_{yy}(\bm{k})\tau_y\sigma_y
    +g_{yz}(\bm{k})\tau_y\sigma_z,
\end{align}
where $\sigma$ and $\tau$ are Pauli matrizes of spin and orbital degrees. If the orbital make-up requires a nontrivial implementation of inversion, some spatial indices will be permuted. Due to the high symmetry in this four-band model, for $g_{x0}=g_{yx}=g_{yy}=0$, the basis can be chosen pairwise such that one can express dispersion and momentum matrix elements referring to only one upper and one lower band index. In the general case one obtains not one but three structurally identical pieces corresponding to the three different interband transitions.
Inserting the degeneracies into the general expression,
and using that $\omega_{nn}=0$, $\omega_{mn}=-\omega_{nm}$, $v_{nm}^a={v_{mn}^a}^*$, one obtains for the non-linear dc conductivity explicitly 
\begin{align}
    \Sigma&^{ab;c}_{\pm}(\omega)\notag\\
    &=\frac{e^3}{\hbar^2\omega^2}
    \int_{\bm{k}}
    \frac{f_{du}}{-i\gamma}
    \left(
    \frac{\Re(v_{du}^a v_{ud}^b) v_{dd}^c}{\varepsilon_{du}-\omega-i\Gamma}
    +\frac{\Re(v_{du}^a v_{ud}^b) v_{uu}^c}{\varepsilon_{du}+\omega+i\Gamma}
    \right)\notag\\
    &+
    \int_{\bm{k}}
    f_{du}
    \left(
    \frac{\Re(v_{du}^a v_{ud}^c)v_{uu}^b }{(\varepsilon_{du}-i\gamma)(\varepsilon_{du}-\omega-i\Gamma)}
    \right.\notag\\&\left.
    -\frac{\Re(v_{du}^a v_{ud}^c)v_{dd}^b }{(\varepsilon_{du}+i\gamma)(\varepsilon_{du}+\omega+i\Gamma)}
    \right)\notag\\
    &\pm(\omega\to -\omega)+\dots 
    \label{eq:ptconductivity}
\end{align}
where the upper (lower) band is labeled $u$ ($d$). 
The dots indicate that there are analogous terms including the other three pairings of bands (involving  the $\mathcal{T}$-partners $\bar u$, $\bar d$), where $v_{ud}\neq v_{\bar u d}\neq v_{u\bar d}\neq v_{\bar u\bar d}$. 
The problem thus decomposes in four pieces which somewhat resembles the two-band case when $\mathcal{T}$ is preserved~\cite{Cook2017}. However, now there are both two-band and three-band terms.

\subsection{Single cone}\label{singlecone}

Concentrating on one band touching point at a time, the local two-band Hamiltonian corresponding to band $(d,u)$ is $H(\bm{k})=\hbar g_i \sigma_i$, with a generic momentum dependencies $g_i=v_F(k_i + \alpha_{ijl}k_jk_l)$; analogous expressions with dissimilar coefficients exist for the other combinations of bands at the remaining Weyl points. 
Keeping only the two-band expression for linear polarization, the injection current becomes
\begin{align}
    \sigma^{aa;c}_{dc}(\omega)
        &=
    -\frac{2e^3}{\hbar^2\omega^2 \gamma}
    \int_{\bm{k}}
    f_{du} |v_{du}^a|^2 (v_{dd}^c-v_{uu}^c)
    \notag\\&\quad\times
    \Im\left(
    \frac{1}{\varepsilon_{du}-\omega-i\Gamma}
    +\frac{1}{\varepsilon_{du}+\omega+i\Gamma}
    \right)
\end{align}
The velocity matrix elements are given by
\begin{align}
    \Re(v^a_{du}v^b_{ud})v^c_{dd}&=
    \hbar^3\sum_{ijl}\partial^a g_{i}\partial^b g_{j}\partial^c g_{l} \frac{g_i g_j g_l}
    {\varepsilon_i\varepsilon_j\varepsilon_l}
     (\delta_{ij}-1).
    \label{eq:gproduct}
\end{align}
Expanding in the non-linearity $\alpha$, the zeroth term vanishes, as expected. At the next order, since $i\neq j$ in Eq~\eqref{eq:gproduct}, the only contribution is ($\Gamma\to 0$)
\begin{align}
    \sigma^{aa;c}_{dc}(\omega)
    &=-\frac{2e^3(v_F\hbar)^3}{\hbar^2\omega^2 \gamma}
    \int_{\bm{k}}
    8\delta_{cl}\delta_{ai}
    \alpha_{iaq} k_q 
    \frac{k_i k_j k_l}{ |\bm{k}|^{3}}
    \notag\\&\quad
    (\delta_{ij}-1)\delta(2v_F|\bm{k}|-\omega)
    \shortintertext{so the two cases of linear polarization are}
    \sigma^{aa;a}_{dc}(\omega)
    &=\frac{16e^3(v_F\hbar)^3}{\hbar^2\omega^2 \gamma}
    \int_{\bm{k}}
    \sum_{i\neq a}\alpha_{iai}\frac{k_a^2k_i^2}{|\bm{k}|^3}\delta(2v_F|\bm{k}|-\omega)
    \notag\\
    &=\frac{2}{3\pi}\frac{e^3}{\hbar^2}
    \sum_{i\neq a}\frac{\alpha_{iai}}{\gamma}\omega
    \\
    \sigma^{aa;c}_{dc}(\omega)
    &=\frac{16e^3(v_F\hbar)^3}{\hbar^2 \omega\gamma}
    \int_{\bm{k}}
    \alpha_{caa}\frac{k_a^2k_c^2}{|\bm{k}|^3}\delta(2v_F|\bm{k}|-\omega)
    \notag\\
    &=\frac{2}{3\pi}\frac{e^3}{\hbar^2}
    \frac{\alpha_{caa}}{\gamma}\omega 
    \qquad\text{for } c\neq a.
    \label{eq:ptexplicit2}
\end{align}
For finite interband relaxation rates ($\Gamma\neq 0$), there is a crossover at frequency $\omega_c\sim (\Gamma \Lambda^2)^{1/3}$, with  $\Lambda$ the bandwidth of the system. Below this $\omega_c$, $\sigma^{aa;c}_{dc}$ acquires a residual term of the size $\sim\Gamma\Lambda^2/\gamma\omega^{2}$.

Barring any symmetry cancellations, for transitions near the Dirac point we thus obtain the estimate
\begin{align}
    \sigma^{aa;c}_{dc}(\omega)&=
   c_1\frac{\Gamma \Lambda^2}{\omega^{2}\gamma}
   +\left(\frac{c_2}{\gamma}+c_3\right)\omega,
\end{align}
where $c_1,c_2,c_3$ are coefficients depending on the non-linearity in the Dirac cone dispersion. The  non-identical Dirac cone close to $-\bm{k}$ will contribute a similar term with numerically different coefficients, but this does not change the the qualitative behavior. Additionally, the values of $c_1$ and $c_3$ are not exclusively determined by diagrams (I) and (II) but will also receive a contribution from the remaining second-order diagrams.

In summary, in case that both $\mathcal{T}$ and inversion are preserved, the second order dc-response will vanish. With $T$ present but $P$ being absent, what matters is the imaginary part of the product of three velocities ($N_{nml}^{abc}$), with $P$ present but without $T$, both the real and imaginary part of $N_{nml}^{abc}$ matters, and in the absence of both $P$ and $T$ but with $\mathcal{T}$-symmetry only the real part of $N_{nml}^{abc}$ remains.
For the $\mathcal{T}$-symmetric case, weak $\mathcal{T}$ breaking will therefore result in a strong suppression of the non-linear dc-conductivity due to the nearly complete cancellation of the contributions from $\bm{k}$ with the ones from $-\bm{k}$. A strong response is therefore expected from intrinsically $\mathcal{T}$-breaking materials where the $\mathcal{T}$ is not broken perturbatively. At leading order the size of the injection current is dependent on the band structure near the Dirac cone through the combination $\alpha/\gamma$, which means that a Dirac cone with a large non-linear components of type $\alpha_{cac}$ or $\alpha_{caa}$ appears to be the most promising candidate for measuring a large dc-current.

Regarding the contributions from diagrams (III)-(VI) which we safely discarded, the situation becomes quite different if $\mathcal{T}$ is preserved, rendering diagrams (I) and (II) of size $\mathcal{O}(1)$ instead of $\mathcal{O}(\gamma^{-1})$. Then, all diagrams have a similar sized contribution to the dc-current~\cite{Cook2017}.

\subsection{Minimal four band model}

We corroborate the findings of the Sec. \ref{singlecone} with an explicit calculation of the BPVE in a $\mathcal{PT}$-symmetric four band model. As a comparison, a similar four-band model exhibiting $\mathcal{T}$ with comparable dispersion and density of states is chosen. They are a representative of the BPVE which arises from two-photon processes in a Dirac cone close to the nodal point. Of course, at higher frequency resonances will appear which are particular to the chosen band structure. To describe the general photo-response beyond the injection current, we should consider all the diagrams in Fig.~\ref{fig:diags}, which corresponds to Eq. \ref{eq:generalsemics}.

\begin{figure}
    \includegraphics[width=.4\columnwidth]{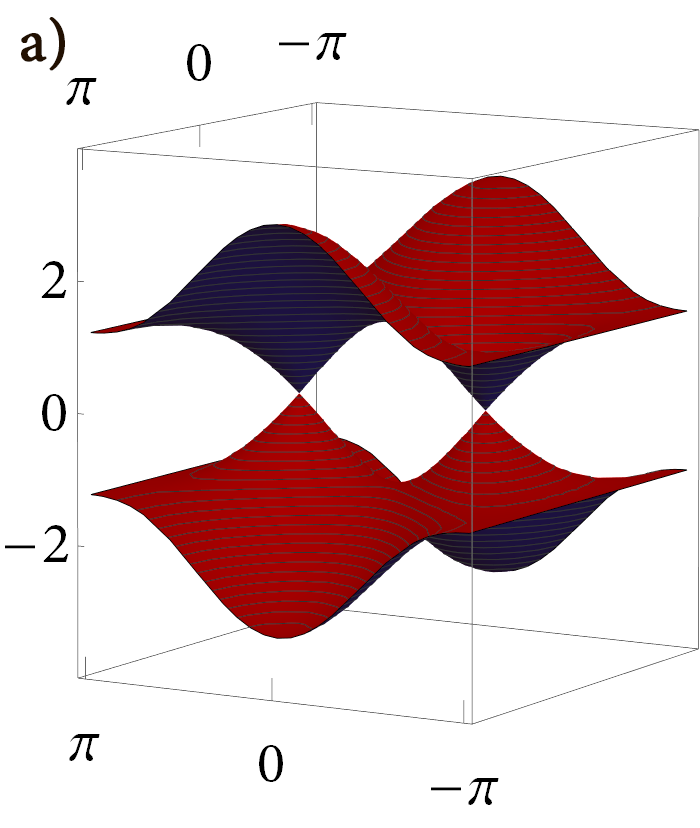}
    \includegraphics[width=.4\columnwidth]{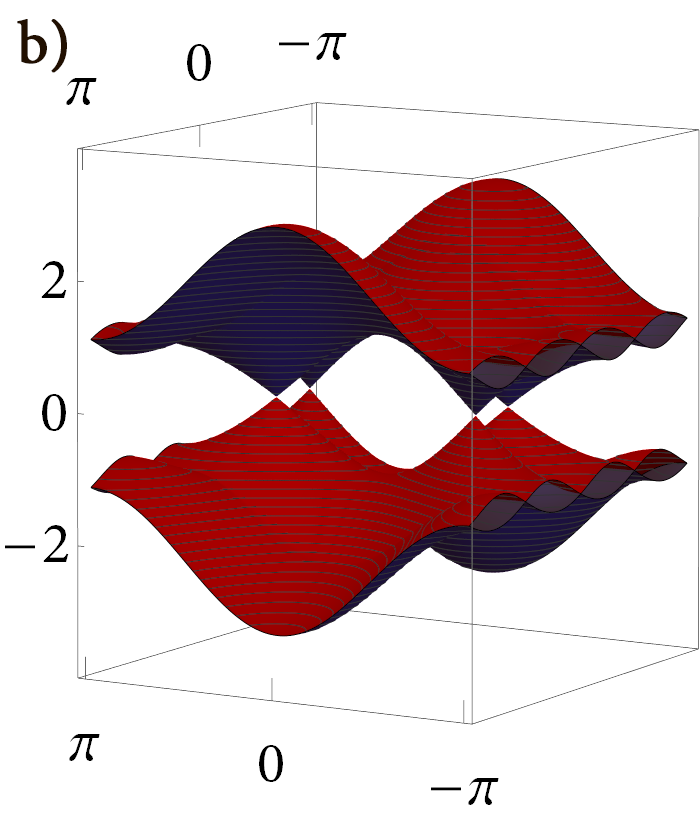}
    \includegraphics[width=.8\columnwidth]{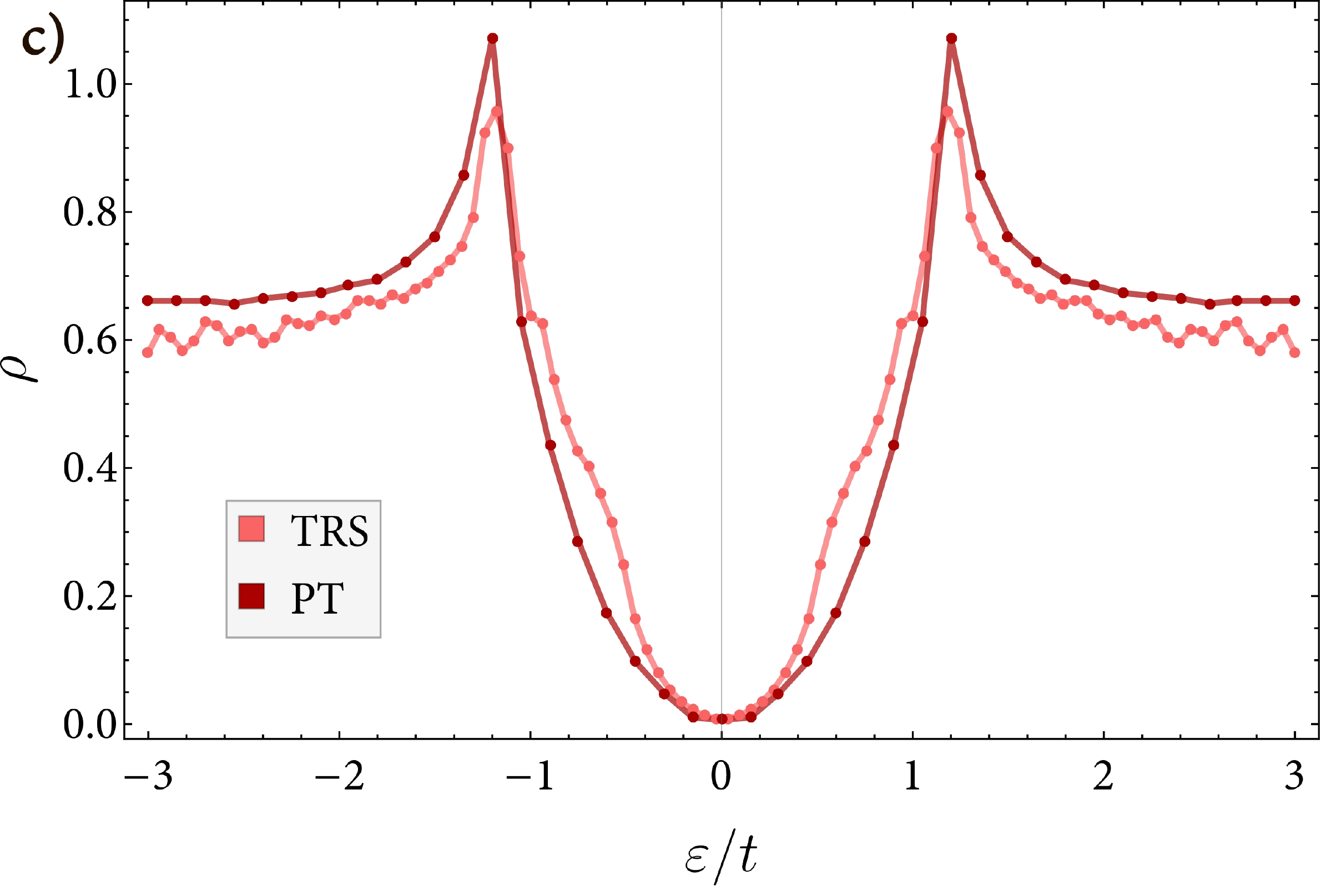}
    \caption{Band structure in the $(k_x,k_y)$-plane at $k_z=0$, for the $\mathcal{PT}$-symmetric Hamiltonian (a) and with the $\mathcal{T}$-symmetric one (b). The parameters are chosen as $m_y=1$, $m_z=5$, $(\Delta,\delta)=(0,0.5)$ (for $\mathcal{PT}$) and $(\Delta,\delta)=(0.5,0)$ (for $\mathcal{T}$). (c) Comparison of the density of states in both cases as a function of chemical potential.}
    \label{fig:3}
\end{figure}

As tight binding Hamiltonian we choose,
\begin{align}
H&=
\hbar t\Bigl\{ 
\sigma_y \sin k_y+  s_x \sigma_z \sin k_z 
\notag\\&\qquad
+\sigma_x s_0 \bigl[\cos k_x + m_y (1 - \cos k_y)
+m_z (1-\cos k_z)\bigr]
\notag\\&\qquad
+\Delta \bigl[s_x\sigma_y \cos k_y+s_z\sin k_z \bigr]
\notag\\ &\qquad
 + \delta \bigl[s_y \sigma_z\cos k_x  + s_z \sigma_z(\cos k_x+\sin k_z)  \bigr]\Bigr\}
 \label{eq:modelHam}
\end{align}
The parameters $m_y, m_z$ produce anisotropy, $\Delta$ breaks inversion symmetry. $\delta$ induces time-reversal and inversion breaking, such that while the Hamiltonian does not commute with either $\mathcal{P}$ or $\mathcal{T}$, their combined operation $\mathcal{PT}$ is a symmetry. The spectrum is then of two doubly-degenerate bands.
In the following we compare the non-linear conductivity between the choice of parameters  $\Delta=0, \delta\neq0$, implementing $\mathcal{T}$ and the choice $\Delta\neq0, \delta=0$ for $\mathcal{PT}$-symmetry.
The band structures have the form shown in Fig. \ref{fig:3}. Both cases lead to a very similar total density of states (DOS) (Fig.~\ref{fig:3}c), with a clear separation between the low energy Dirac cone physics and the high energy Lifshitz point where the cones connect. 
\begin{figure}
    \includegraphics[width=\columnwidth]{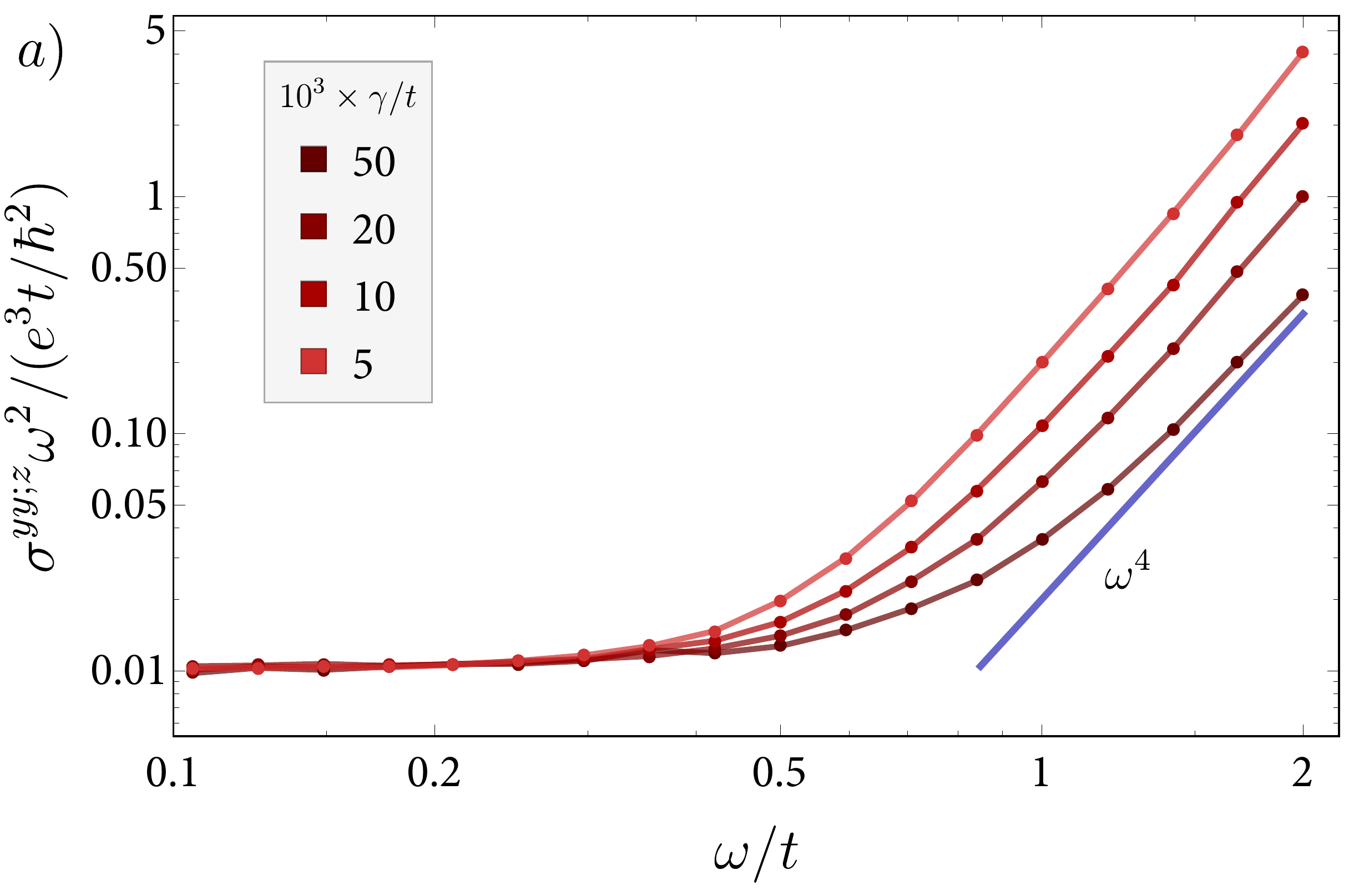}
    \includegraphics[width=\columnwidth]{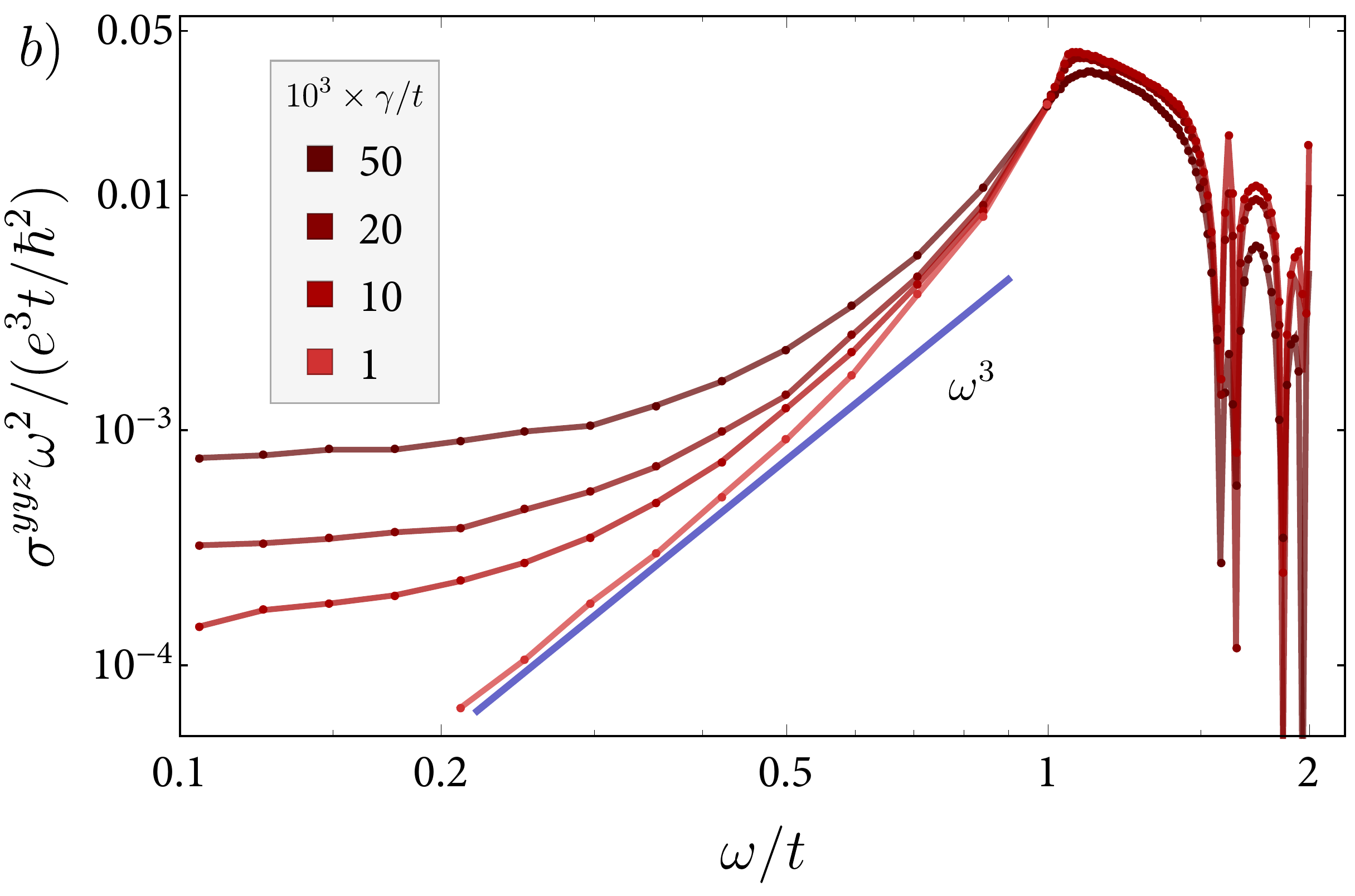}
    \caption{Comparison of the modulus of $\sigma^{yy;z}(0,\omega,-\omega)$ between the $\mathcal{PT}$-symmetric case (a) and the one with $\mathcal{T}$ (b). For improved visibility, we show the product $\sigma^{yy;z}\omega^2$. For $\mathcal{PT}$, the non-linear dc-conductivity approaches a $\omega^{-2}$-dependence for small frequencies which is independent of the quasiparticle lifetime $\gamma$. At frequencies above a threshold proportional to $\Gamma$, the current creation is only capped by the lifetime $1/\gamma$.
    For $\mathcal{T}$, the current vanishes linearly with $\omega^1$ for small frequencies, save for a residual term of size $\gamma\omega^{-2}$ for the smallest frequencies. At frequencies comparable to the bandwidth, the spectrum is sensitive to nesting effects, and thus to the band structure. Note that the apparent erratic behavior for these large values of $\omega$ is due to the logarithmic scale, not numerical artefacts.
    Parameter values are the same as in Fig.~\ref{fig:3}.
    }
    \label{fig:5}
\end{figure}

The Hamiltonian, Eq.~\eqref{eq:modelHam} is generic in the sense that it encodes the minimal number of four Weyl cones in case of $\mathcal{T}$  and the minimal number of two doubly degenerate Dirac cones in the $\mathcal{PT}$ case. The momentum dependencies are not generic, they are chosen in reference to a minimal model which was proposed for the time-reversal symmetric material TaAs, which has a cubic Brillouin zone. In Sec.~\ref{app:B}, the steps are listed which lead to Eq.~\eqref{eq:modelHam}, suffice to say here that the only nontrivial part is to keep enough structure in the momentum dependencies such that there are no unwanted additional symmetries like mirrors present. There is one specialty added to the $\mathcal{PT}$-symmetric dispersion which is to not include a term which would correspond to a non-linearity in the dispersion of the type $\alpha_{zyy}$ in the vicinity of the Dirac points. This allows us to verify the claim of  Eq.~\eqref{eq:ptexplicit2} that the injection current is sourced from exactly one non-linear dispersion term at next-to-leading order. In other words, the expected injection current for our $\mathcal{PT}$-symmetric example for the dc-conductivity $\sigma^{yy;z}$ has size $\mathcal{O}(\omega^{2}/\gamma)$ instead of $\mathcal{O}(\omega/\gamma)$. For the numerical evaluation we repeat that diagrams (III)-(VI) are negligibly small, so that it is enough to consider diagrams (I) and (II), with varying but finite relaxation rates $\gamma=\Gamma$. The second order dc-conductivity $\sigma^{yy;z}$ is shown in Fig.~\ref{fig:5} as a function of frequency. As expected, the injection current persists for large frequencies, crossing over into a shift current for small $\omega$. 
We emphasize the difference in magnitude of roughly $10^2$ between $\mathcal{T}$ and $\mathcal{PT}$, in spite of the very comparable band structure, bandwidth, and relaxation rates used for both models.

\section{Conclusions}
For many years, studies of the bulk photovoltaic effect were restricted to the evaluation of the complicated perturbative expressions. 
Here, we used the diagrammatic approach to describe second-order optical response for materials without time-reversal symmetry and also with finite relaxation times in terms of four distinct physical processes. 
The central element was the analysis of the anomalous acceleration, which appears naturally in the velocity gauge.
The resulting analytical structure (e.g. Eqs. \ref{eq:generalsemics}, \ref{eq:correctinj} and \ref{eq:correctshift}) of the expressions is substantially more amenable to analytical manipulations, but can also be directly implemented in numerics.

We derived explicit sum rules for non-quadratic Hamiltonians and demonstrated that their cancellation properties are equivalently accounted for in the diagrammatic formulation of the velocity gauge. 

Contrary to the semiclassical intuition, the presence of a Berry dipole is not a necessary requirement for a large BPVE.
Instead, the topological content of the second-order response is encoded in expressions which involve both the antisymmetric and the symmetric part of the quantum geometric tensor. 
In particular, we showed that the shift current can be written as the sum of a piece containing the quantum metric and certain elements of the Berry dipole. The shift current is therefore of topological origin, but not in the usual sense that it would be proportional to the Berry curvature. Also, it does not evaluate to a quantized value in a topological insulator, 
and it contains a non-universal piece which we connected to the ballistic current. In agreement with earlier work, we recover the Berry dipole in the semiclassical expansion, but alongside find a number of other terms which only vanish in the presence of $\mathcal{T}$. This means that the semiclassical current response is of topological origin only in time-reversal symmetric compounds. From the same analysis, we conclude that the injection current is generically not of topological origin, instead being attributed to \emph{dynamical antilocalization} between counter-propagating momentum space trajectories. 

We believe that the same analysis is feasible for third order conductivities, with a larger number of physically distinct processes emerging for the matrix elements involving third derivatives ('jerk').

The principles outlined here should prove helpful in the search for materials with tailor-made non-linear optical response. For example, we outlined that current injection is enhanced in semimetals with broken time-reversal symmetry when they exhibit a large non-linear term in their dispersion near the Dirac cone.

\begin{acknowledgments}
We thank 
S. Fischer,
O. Golan,
J. Hofmann,
R. Thorngren,
A. Stern,
M. Mittnenzweig,
Y. Zhang,
H. Ishizuka,
N. Nagaosa,
L. Wu,
F. de Juan Sanz, 
A. G. Grushin, and
D. E. Parker
for helpful discussion. 
B.Y. acknowledges the financial support by the Willner Family Leadership Institute for the Weizmann Institute of Science, the Benoziyo Endowment Fund for the Advancement of Science,  Ruth and Herman Albert Scholars Program for New Scientists, the European Research Council (ERC) under the European Union's Horizon 2020 research and innovation programme (Grant No. 815869).
\end{acknowledgments}

\appendix

\section{Relaxation rates}
\label{app:relax}
Here we give some further details on the nature of the finite relaxation rates $\gamma$ and $\Gamma$ in the three-leg diagrams.
In a Fermi liquid, lifetime effects enter the propagator through the retarded self-energy $\Sigma^R(\bm{k},\omega)$, according to
$\Sigma^R_m(\bm{k},\omega)={\mathcal{G}^R_m}^{-1}(\bm{k},\omega)-{G_m^R}^{-1}(\bm{k},\omega)$.
where ${G^R_m}^{-1}(\bm{k},\omega)=\omega+i0^+-\varepsilon_m(\bm{k})$ is the bare retarded Green's function. In the the relaxation time approximation, the self-energy is projected onto the Fermi surface and averaged, thus replacing it by a constant. This procedure is ubiquitously used for ground state calculations. 
Due to the multiband nature of the optical response, here the reference point is not always the Fermi surface but the respective band where the electron is travelling, i.e. 
$\langle \Sigma^R_m(\bm{k},\omega)\rangle_{\bm{k}}=i/2\tau_{m}$. After the contour integration in the loop frequency $\Omega$, the triple product of propagators $\mathcal{G}^R(\bm{k},\Omega)
\mathcal{G}^{R/A}(\bm{k},\Omega+\omega_1)
\mathcal{G}^A(\bm{k},\Omega+\omega_1+\omega_2)$ leads to several terms with two denominators. 
In these, three frequency combinations appear, $\omega_1$, $\omega_2$ and $\omega_1+\omega_2$, each of them with the imaginary part $1/2\tau_m+1/2\tau_n$. The only question is which combinations of band indices are the relevant ones for the zeros of the real part of the  denominators.

If we consider the case that $\omega_1=-\omega_2$, two Greens functions are at the same energy, as shown in Fig.~\ref{fig:onshell} in the main text.
The denominators with $\omega_{1,2}$ have a pole at $\omega_{1,2}=\varepsilon_{mn}$, where one band index is occupied, and one is unoccupied. This leads to the interband relaxation rate $\Gamma=1/2\tau_c+1/2\tau_v$, where the conduction band with index $c$ and the valence band with index $v$ are chosen so that  $|\omega_{1,2}|=\varepsilon_{cv}$. Other combination of band indices have different interband relaxation rates, but they are not resonant, so that it is always true that $|\omega_{1,2}-\varepsilon_{mn}|\gg \Gamma$. For this reason, in a given denominator, it is enough to specify one relaxation rate. For the combination $\omega_1+\omega_2=0$, the pole is realized for $\varepsilon_{mn}=0$, which means that the process is either intraband or the bands $m$ and $n$ are degenerate. In both cases, the relaxation rate is given by $\gamma=1/2\tau_v+1/2\tau_v$, which only depends on the microscopic quasiparticle lifetime in the valence band(s). Again, the other combination of band indices in this specific denominator are insensitive to the precise value of the relaxation rate, as it holds again that $|\omega_{1}+\omega_2-\varepsilon_{mn}|\gg \gamma$.

\section{Derivatives of the Berry connection}
\label{app:A}
In the following, we establish the general procedure how to easily derive the sum rules necessary for a translation between length gauge and velocity gauge. 
The first step is to expand derivatives of the matrix elements $r^a_{mn}$ into commutators. To our knowledge this has not been documented so far.
We use the definition of the momentum space derivative for a multiband system,
\begin{align}
    \langle m| (\partial^a\mathcal{O})|n \rangle
    &=\partial^a \mathcal{O}_{mn}
    -i r^a_{ml}\mathcal{O}_{ln}+i \mathcal{O}_{ml}r^a_{ln}
\end{align}
This represents a known generalization of the usual Berry connection. Note that this non-Abelian connection is covariant under gauge transformations, not invariant. The momentum derivative on the left hand side can be evaluated directly as long as the operator is given in a basis which is momentum independent, which applies in particular to a tight-binding Hamiltonian. However, for $r^a_{mn}$ the left hand side is of little use as the derivative cannot act on anything except the wavefunction, which means we do not know how to write the operator in any other basis than the canonical band basis. Therefore, we only evaluate the right hand side and use this to define the matrix elements on the left hand side.
An important role is played by the  second derivative $\langle m|i\partial^b\partial^c n\rangle 
-\langle i\partial^b\partial^c  m|n\rangle
=2R^{bc}_{mn}=2R^{cb}_{mn}$.
One can  equivalently rewrite this by using
%\begin{widetext}
\begin{align}
	\partial^b\partial^c(\langle m| n\rangle)=0
	&=
	\langle \partial^b m| \partial^c n \rangle
	+\langle \partial^c m| \partial^b n \rangle
	\notag\\&\quad
	+\langle m|\partial^b\partial^c n\rangle
	+\langle \partial^b\partial^c m| n\rangle\\
	\langle m|\partial^b\partial^c n\rangle
	+\langle \partial^b\partial^c m| n\rangle
	&=\langle \partial^b m| \partial^c n \rangle
	+\langle \partial^c m| \partial^b n \rangle.
\end{align} 
We further observe that
\begin{align}
	\langle m|\partial^b\partial^c n\rangle
	&=
	\langle m|\partial^b\Bigl(\sum_l|l \rangle 
	\langle l|\partial^c n\rangle\Bigr)\\
	&=\sum_l\langle m|\partial^b l \rangle 
	\langle l|\partial^c n\rangle
	+\langle \partial^b m| \partial^c n \rangle
	+\langle m|\partial^b\partial^c n\rangle\\
	\langle \partial^b m| \partial^c n \rangle
	&=\sum_l r^b_{ml}r^c_{ln}
\end{align}
In other words, the quantity $\langle m|\partial^b\partial^c n\rangle+\langle \partial^b\partial^c m| n\rangle$ is essentially the symmetrized matrix element $\langle \partial^b m| \partial^c n \rangle$, a quantity known as the (bare) quantum metric~\cite{Cheng2010}. 
Importantly, the diagonal matrix elements of the bare quantum metric are not gauge invariant. The proper gauge invariant distance measure is instead introduced via the quantum geometric tensor $Q^{ab}_{mm}$.
The quantum geometric tensor is defined with the help of the projector $\mathcal{P}_{GS}=\sum_l^{N} |l \rangle 
	\langle l|$ onto the $N$-dimensional ground state in band basis, and reads~\cite{Ma2010}
\begin{align}
    Q^{bc}_{mn}&=\langle \partial^b m|1-\mathcal{P}_{GS}|\partial^c n\rangle\\
    &=\sum_l r^b_{ml}r^c_{ln}-
    \sum_l^N r^b_{ml}r^c_{ln}.
\end{align}
Here, we used that $\langle \partial^b m|n \rangle= -\langle m| \partial^b n \rangle$.
The non-Abelian quantum metric $g^{ab}$ is the symmetric part of $Q^{ab}$, while the non-Abelian Berry curvature is given by the antisymmetric part, i.e. $Q^{ab}_{mn}=g^{ab}_{mn}-\tfrac{i}{2}\Omega^{ab}_{mn}$. Therefore, by definition the matrix elements $2iR^{ab}_{mn}=\langle  \partial_{k_a}\partial_{k_b} m| n\rangle-\langle m|  \partial_{k_a}\partial_{k_b} n\rangle$ are related to the quantum metric by
\begin{align}
	2g^{ab}_{mm}&=
	2iR^{ab}_{mm}
+2\langle m| \partial^a\partial^b m\rangle
\notag\\&\quad
	-\sum_l f_l (r^b_{ml}r^c_{lm}
	+ r^c_{ml}r^b_{lm}),
\end{align}
where we used the Fermi factor $f_l$ to perform the sum over the ground state manifold. One might want to go about and immediately construct the response in terms of the gauge-invariant quantities $g$ and $\Omega$. However, as we show in the main text, this is not recommendable. Instead, the matrix elements which appear in the conductivity are, by construction, always diagonal matrix elements of commutators of Hermitian operators. As such, while the constituents are not gauge-invariant, the end result is. The same observation was reached previously on the basis of general properties of the fiber bundle in Bloch basis~\cite{Parker2019}.

We decompose the momentum derivative in symmetric and antisymmetric parts, yielding $R^{bc}=\partial^b r^c-\tfrac{i}{2}[r^b,r^c]$. Equivalently, this means,
\begin{align}
    \partial^b r^c-\partial^c r^b&=i[r^b,r^c]\\
    2R^{bc}&=\partial^b r^c+\partial^c r^b.
\end{align}
As usual, the Berry curvature is $\Omega^{bc}=\partial^b r^c-\partial^c r^b=i[r^b,r^c]$.
Repeating the same procedure for the Berry curvature dipole results in the diagonal components
\begin{align}
    \partial^a\Omega^{bc}&=-\tfrac{1}{2}[r^a, [r^b,r^c]]+i[r^b,R^{ac}] - i[r^c,R^{ab}],
\end{align}
where we used the identity $0 =[r^b, [r^c,r^a]]+[r^c,[r^a,r^b]]+[r^a,[r^b,r^c]]$.
The second step involves manipulations of objects containing velocity matrix elements $v^a_{mn}$ and their derivatives. We use the relation that for off-diagonal matrix elements  $\partial^av^b=i\varepsilon\partial^ar^b+i\Delta^av^b+\mathcal{O}(\Gamma)$ and, as previously established, $\partial^b r^c=R^{bc}+\tfrac{i}{2}[r^b,r^c]$.
For example, for the operator $\partial^a\partial^b H_0(\bm{k})$ this yields immediately
\begin{align}
    w^{ab}_{mn}&=\partial^av^b-i[r^a,v^b]
    \qquad m\neq n\notag\\
    &=i\varepsilon(R^{ab}+\tfrac{i}{2}[r^a,r^b])+i\Delta^ar^b-i[r^a,v^b]+\mathcal{O}(\Gamma).
\end{align}
Under the reasonable assumption that the dispersion relation is mostly smooth, it is $w^{ab}=w^{ba}$. In the main text, this was used to write the result more concisely. However, it also means that
\begin{align}
    -\tfrac{\varepsilon}{2}[r^a,r^b]&+i\Delta^ar^b
    -i[r^a,v^b]
    \notag\\&=
    -\tfrac{\varepsilon}{2}[r^b,r^a]+i\Delta^br^a
    -i[r^b,v^a]\\
    i\varepsilon [r^a,r^b]&=
    [r^a,v^b]+[v^a,r^b]
    -\Delta^a r^b+\Delta^b r^a.
\end{align}
Therefore, also
\begin{align}
    -[v^c,[r^a,r^b]]&=
    [r^c,[r^a,v^b]]+[r^c,[v^a,r^b]]
    \notag\\&\quad
    -[r^c,\Delta^a r^b]+[r^c,\Delta^b r^a]
    \label{eq:app1}
\end{align}
In the same manner, we conclude that
\begin{align}
    u^{abc}&=
    \partial^a\partial^bv^c
    -i[\partial^a r^b,v^c]
    -i[r^b,\partial^a v^c]
    \notag\\&\quad
    -i[ r^a,\partial^b v^c]
    -[ r^a,[r^b,v^c]]
    \qquad (m=n)
    \notag\\
    &=\partial^a\partial^bv^c
    -[r^a,[r^b,v^c]]
    +[r^b,\Delta^a r^c]
    +[r^a,\Delta^b r^c]
    \notag\\&\quad
    -\tfrac{1}{2}[v^c,[r^a,r^b]]
    -\tfrac{1}{2}[v^a,[r^b,r^c]]
    +\tfrac{1}{2}[v^b,[r^c,r^a]]
    \notag\\&\quad
    +i[v^c,R^{ab}]+i[v^b,R^{ca}]+i[v^a,R^{bc}]
    +\mathcal{O}(\Gamma).
\end{align}
By using that $[r^a,[r^b,v^c]]+[r^b,[v^c,r^a]]+[v^c,[r^a,r^b]]=0$, 
this becomes
\begin{align}
    u^{abc}&=
    \partial^a\partial^bv^c
    -[r^a,[r^b,v^c]]
    +[r^b,\Delta^a r^c]+[r^a,\Delta^b r^c]
    \notag\\&\quad
    +i[v^c,R^{ab}]+i[v^b,R^{ca}]+i[v^a,R^{bc}]
\end{align}
Additionally it holds that
\begin{align}
    u^{abc}-u^{bac}&=0=
    -[r^b,[v^c,r^a]]+[r^a,[v^c,r^b]]\\
    [v^c,[r^a,r^b]]&=[r^c,[r^a,v^b]]+[r^c,[v^a,r^b]]
    \notag\\&\quad
    -[r^c,\Delta^a r^b]+[r^c,\Delta^b r^a]=0.
\end{align}
We emphasize that we do not need all of these relations, but they allow for the sum rules to be rearranged in a condensed format of highly symmetric or cyclical expressions using only velocity matrix elements and wave function derivatives of increasingly higher order depending on the order considered for the non-linear conductivity. In the main text, we show how these various terms can be interpreted physically.

\section{Comments on the four-band model}
\label{app:B}
As a stand-in for a generic four band model we make use of the tight-binding Hamiltonian previously proposed for TaAs, which reads with normalized hopping
\begin{align}
    H&=[\cos k_x
    +m_y(1-\cos k_y)
    +m_z(1-\cos k_z)] \sigma_x
    \notag\\&
    +\sin k_y\sigma_y+\Delta \cos k_y \sigma_y s_x+
    \sin k_z \sigma_z s_x.
\end{align}
Here, $\sigma$ are orbital degrees of freedom and $s$ spin. $\Delta$ breaks inversion. Denoting complex conjugation by $\mathcal{K}$ and the replacement $\bm{k}\rightarrow -\bm{k}$ by $\mathcal{M}$ , it holds generally for the time reversal operator in momentum space
\begin{align}
    \mathcal{T}&=i s_y \mathcal{K}\mathcal{M}
\end{align}
From the Hamiltonian we further deduce the appropriate form of the inversion operator to
\begin{align}
    \mathcal{P}&=\sigma_x \mathcal{M}.
\end{align}
This is because just by inversion we flip the signs of the $\sin$-terms, which can be rectified by flipping exactly $\sigma_y$ and $\sigma_z$, something achieved by $\sigma_x$. Other models will have different $\mathcal{P}$-operators.

The original model also has two mirror symmetries, implemented by
\begin{align}
    \mathcal{X}_x&=\mathcal{M}_x s_x\\
    \mathcal{X}_y&=\mathcal{M}_y \sigma_x s_y.
\end{align}
This is undesirable because the mirror planes can lead to additional cancellations which might skew the result against what we are trying t demonstrate, i.e. that a $\mathcal{PT}$-symmetric material can have larger response than a otherwise very similar $\mathcal{T}$ material.
For this reason, we break the mirrors by adding  $\Delta \sin k_z s_z$ to the time-reversal symmetric Hamiltonian.

The operator for $\mathcal{PT}$ reads $\mathcal{PT}=i\sigma_x s_y \mathcal{K}$. 
This means that altogether there are the following five elements in (pseudo-)spin space which preserve this combined symmetry, $\mathcal{PT}$
\begin{align}
    \sigma_x\quad
    \sigma_y\quad
    \sigma_z s_x\quad
    \sigma_z s_y\quad   
    \sigma_z s_z
\end{align}
All of them can also preserve $\mathcal{P}$ and $\mathcal{T}$ separately given the corresponding sign change is compensated by an appropriate momentum dependence.

In close analogy with the $\mathcal{T}$ case, for the $\mathcal{PT}$-symmetric Hamiltonian we thus choose
\begin{align}
    H&=[\cos k_x
    +m_y(1-\cos k_y)
    +m_z(1-\cos k_z)] \sigma_x
    \notag\\&
    +\sin k_y\sigma_y
    +\sin k_z \sigma_z s_x
    \notag\\&
    +\delta[\cos k_x s_y \sigma_z + (\cos k_x+\sin k_z) s_z \sigma_z],
\end{align}
where the momentum dependence of the last term allows for a Dirac point along the x-axis.

%\bibliography{literature}

\clearpage

\end{document}